\documentclass{aa}  
\usepackage{graphicx}
\catcode`\@=11
\def\gtsima{\ifmmode{\mathrel{\mathpalette\@versim>}}
    \else{$\mathrel{\mathpalette\@versim>}$}\fi}
\def\ltsima{\ifmmode{\mathrel{\mathpalette\@versim<}}
    \else{$\mathrel{\mathpalette\@versim<}$}\fi}
\def\@versim#1#2{\lower 2.9truept \vbox{\baselineskip 0pt \lineskip 
    0.5truept \ialign{$\m@th#1\hfil##\hfil$\crcr#2\crcr\sim\crcr}}}
\catcode`\@=12
\def\lae{Ly$\alpha$ }
\usepackage{txfonts}
\begin{document}
   \title{The VIMOS VLT Deep Survey: Star formation rate density of
          Ly$\alpha$ emitters from a sample of 217 galaxies with
          spectroscopic redshifts $2 \leq z \leq 6.6$\thanks{Based on
          data obtained with the European Southern Observatory Very
          Large Telescope, Paranal, Chile, under Large Programs
          070.A-9007 and 177.A-0837.  Based on observations obtained
          with MegaPrime/MegaCam, a joint project of CFHT and
          CEA/DAPNIA, at the Canada-France-Hawaii Telescope (CFHT)
          which is operated by the National Research Council (NRC) of
          Canada, the Institut National des Sciences de l'Univers of
          the Centre National de la Recherche Scientifique (CNRS) of
          France, and the University of Hawaii. This work is based in
          part on data products produced at TERAPIX and the Canadian
          Astronomy Data Centre as part of the Canada-France-Hawaii
          Telescope Legacy Survey, a collaborative project of the NRC
          and CNRS.}}  \titlerunning{SFRD from Ly$\alpha$ emitters in
          VVDS}

   \author{P. Cassata \inst{1}, 
O. Le F\`evre \inst{2}, 
B. Garilli \inst{3},
D. Maccagni \inst{3},
V. Le Brun \inst{2},
M. Scodeggio\inst{3},
L. Tresse \inst{2},
O. Ilbert \inst{2},
G. Zamorani \inst{4},
O. Cucciati \inst{2},
T. Contini\inst{5},
R. Bielby\inst{6},
Y. Mellier\inst{6},
H. J. McCracken\inst{6},
A. Pollo\inst{7},
A. Zanichelli\inst{8},
S. Bardelli\inst{4},
A. Cappi\inst{4},
L. Pozzetti\inst{4},
D. Vergani \inst{4},
E. Zucca\inst{4}
}
\authorrunning{P. Cassata et al.}  \offprints{P. Cassata}

   \institute{Department of Astronomy, University of Massachusetts,
     Amherst, MA 01003, USA
              \email{paolo@astro.umass.edu}
   \and
     Laboratoire d'Astrophysique de Marseille, UMR6110,
     CNRS-Universit\'e de Provence Aix-Marseille I, 
     38, rue Fr\'eric Joliot-Curie,
     F-13388 Marseille cedex 13, France
   \and
     IASF-INAF - via Bassini 15, I-20133, Milano, Italy
   \and
     INAF-Osservatorio Astronomico di Bologna - via Ranzani 1, I-40127, Bologna, Italy
     \and
Laboratoire d'Astrophysique de Toulouse-Tarbes, Universit\'e de Toulouse, CNRS, 14 Av. E. Belin, 31400 France
\and
Institut d'Astrophysique de Paris, UMR7095 CNRS, Universit\'e Pierre et Marie Curie, 98 bis Boulevard Arago, 75014 Paris, France 
\and
The Andrzej Soltan Institute for Nuclear Studies, ul. Hoza 69, 00-681 Warszawa, Poland
\and
IRA-INAF, Via Gobetti, 101, I-40129 Bologna, Italy
}

   \date{Received .....; accepted .....}

  \abstract 
{} 
  {The aim of this work is to study the contribution of the
  \lae emitters to the star formation rate density (SFRD) of the Universe
  in the interval $2<z<6.6$.}
  {We assembled a sample of 217 \lae emitters (LAE) from the Vimos-VLT
    Deep Survey (VVDS) with secure spectroscopic redshifts in the
    redshift range $2<z<6.62$ and fluxes down to F$\sim
    1.5\times10^{-18}erg/s/cm^{2}$. 133 \lae emitters are
    serendipitous identifications in the 22 arcmin$^2$ total slit area
    surveyed with the VVDS-Deep and the 3.3 arcmin$^2$ from the VVDS
    Ultra-Deep survey, and 84 are targeted identifications in the 0.62
    deg$^2$ surveyed with the VVDS-DEEP and 0.16 deg$^2$ from the
    Ultra-Deep survey. Among the serendipitous targets we estimate
    that 90\% of the emission lines are most probably Ly$\alpha$,
    while the remaining 10\% could be either [OII]3727 or
    Ly$\alpha$. We omputed the luminosity function and derived the
    star formation density from LAE at these redshifts.}
  {The VVDS--LAE sample reaches faint line fluxes
   $F(Ly\alpha)=1.5\times10^{-18} erg/s/cm^2$ (corresponding to
   $L(Ly\alpha)\sim10^{41} erg/s$ at $z\sim3$) enabling the faint end
   slope of the luminosity function to be constrained to
   $\alpha\sim-1.6 \pm 0.12$ at redshift $z \sim 2.5$ and to
   $\alpha\sim-1.78^{0.10}_{-0.12}$ at redshift $\sim 4$, placing on
   firm statistical grounds trends found in previous LAE studies, and
   indicating that sub-L$_*$ LAE (L$_{Ly-\alpha}\lesssim10^{42.5}
   erg/s$) contribute significantly to the SFRD. The projected number
   density and volume density of faint LAE in $2\leq z \leq 6.6$ with
   F$> 1.5\times10^{-18}erg/s/cm^{2}$ are $33$ galaxies/arcmin$^2$ and
   $\sim 4 \times 10^{-2}$Mpc$^{-3}$, respectively.  We find that the
   the observed luminosity function of LAE does not evolve from z=2 to
   z=6. This implies that, after correction for the redshift-dependent
   IGM absorption, the intrinsic LF must have evolved significantly
   over 3 Gyr. The SFRD from LAE is found to be contributing about
   20\% of the SFRD at $z=2-3$, while the LAE appear to be the dominant
   source of star formation producing ionizing photons in the early
   universe $z\sim>5-6$, becoming equivalent to that of Lyman Break
   galaxies.  }
{}

   \keywords{Cosmology: observations -- Galaxies: fundamental
     parameters -- Galaxies: evolution -- Galaxies: formation }

   \maketitle
%

\section{Introduction}
The \lae line is the strongest hydrogen emission line in the Universe,
and it is observed in the optical range for galaxies at $z>2$. It has
thus naturally been used to search for high-$z$ galaxies
(Partridge\&Peebles~1967; Djorgovski~et~al.~1985; Cowie~\&Hu~1998).

The \lae emission in galaxies is thought to be produced by star
formation, as the AGN contribution to the \lae population at $z<4$ is
found to be less than 5\% (Gawiser~et~al.~2006; Ouchi~et~al.~2008;
Nilsson~et~al.~2009). However, the physical interpretation of the
observed \lae flux is not simple, because the \lae photons are
resonantly scattered by neutral hydrogen. \lae photons can therefore
be more attenuated than other UV photons, and they have an escape
fraction that can depend on the spatial distribution of neutral and
ionized gas, as well as on the velocity field of the neutral gas
(Giavalisco~et~al.~1996; Kunth~et~al.~1998; Mas-Hesse~et~al.~2003;
Deharveng~et~al.~2008 and references therein).

\lae emitters (LAE) observed up to now are forming stars at rates of
$\sim1\div10 M_{\odot}yr^{-1}$ (Cowie\&Hu~1998; Gawiser~et~al.~2006;
Pirzkal~et~al.~2007), and they have stellar masses as low as
$10^8\div10^9 M_{\odot}$ and ages $<50Myr$ (Pirzkal~et~al.~2007; ;
Gawiser~et~al.~2007; Nilsson~et~al.~2009). However,
Nilsson~et~al.~(2007) find ages between 0.1 and 0.9 Gyr, and more
recently Pentericci~et~al.~(2007) and Finkelstein~et~al.~(2009)
point out that \lae galaxies are a more heterogeneous family than
young star-forming galaxies: they also find \lae emitters with old
stellar populations (ages of $\sim1~Gyr$) and a wide range of stellar
masses (up to $10^{10} M_{\odot}$).

Interestingly, a class of \lae galaxies with rest-frame equivalent
width $EW>240$\AA~ has been found (Malhotra\&Roads~2002;
Shimasaku~et~al.~2006). Galaxies with such large EW cannot be
explained by star formation with a Salpeter IMF, but must have a top
heavy IMF, a very young age $<10^7 yr$ and/or a very low metallicity.
Many of these large EW objects are spatially extended, and thus
are good candidates to be cooling clouds or primeval galaxies
(Yang~et~al.~2006; Schaerer~2002) and can give interesting clues about
the first stages of star formation.

Together with studying the properties of \lae galaxies at different
redshifts, it is important to study the evolution of their luminosity
function, comparing large and complete samples of \lae galaxies at
different redshifts.  The most common technique used so far has been
to build large samples from imaging in narrow band filters tuned to
detect \lae emission at $z\sim2\div9$ (Hu~et~al.~2004;
Cuby~et~al.~2003; Tapken~et~al.~2006; Kashikawa~et~al.~2006;
Gronwall~et~al.~2007; Murayama~et~al.~2007; Ouchi~et~al.~2008;
Nilsson~et~al.~2009; Guaita~et~al.~2010). Blank field spectroscopy
has been used blindly to search for \lae emitters in deep HST-ACS
slitless spectroscopic observations (Malhotra~et~al., 2005) or slit
spectroscopy (van~Breukelen,~Jarvis~\&~Venemans~2005; Martin~et~al.~2008;
Rauch~et~al.~2008; Sawicki~et~al.~2008), with Rauch~et~al. (2008)
exploring the faintest emitters.

The general consensus today is that the apparent luminosity function
of \lae galaxies, that is the non-IGM corrected luminosity function,
does not evolve at $z\sim3\div6$ (Rhoads\&Malhotra~2001;
Ouchi~et~al.~2003; van~Breukelen,~Jarvis~\&~Venemans~2005;
Shimasaku~et~al.~2006; Murayama~et~al.~2007; Gronwall~et~al.~2007;
Ouchi~et~al.~2008; Grove~et~al.~2009). However, this conclusion is
drawn from small samples, as the narrow band imaging techniques sample
only thin slices in redshift (typically $\Delta z$=0.1) and needs to
be spectroscopically confirmed. Extensive spectroscopic follow-ups of
narrow band \lae candidates have been carried out in recent years,
gathering hundreds of spectroscopic confirmations between $z\sim2$ and
$z\sim7$. However, the spectroscopic coverage rarely reaches
$30\div50$\% of the photometric sample (Murayama~et~al.~2007;
Gronwall~et~al.~2007; Ouchi~et~al.~2008), but usually is much lower
(Kashikawa~et~al.~2006; Nilsson~et~al.~2007;
Matsuda~et~al.~2005). Moreover, current blind spectroscopic surveys
sample only small areas to relatively shallow fluxes. It is likely
that the apparent lack of evolution is a coincidence of the evolving
intrinsic \lae LF combined with an evolution of the intergalactic
medium absorption with redshift (e.g. Ouchi~et~al.~2008). Firm
conclusions about the possible evolution of the luminosity function
have not yet been secured.  Moreover, existing spectroscopic and
narrow band samples are not sufficiently deep to constrain, even at
intermediate redshift ($z\sim3$), the slope of the luminosity
function.

Measuring the luminosity function in turn enables to compute the
luminosity density and star formation rate density evolution under a
set of well constrained hypotheses. The contribution of \lae to the
total star formation rate is yet not robustly measured mainly because
the faint end slope of the luminosity function remains poorly
constrained.

In this paper, we present the results from a very deep blind
spectroscopic survey search of Ly$\alpha$ emitters over an
unprecedented large sky area. We looked for the serendipitous
detection of \lae emission in the slits of the VIMOS VLT Deep Survey,
concentrating on the VVDS--Deep and VVDS--Ultra-Deep surveys reaching
up to 18h of integration on the VLT-VIMOS. We describe the
spectroscopic and photometric data and the associated selection
function in Section 2. The search for \lae emitters and the final
sample are presented in Section 3, and we discuss its properties in
Section 4. The luminosity function calculation is discussed in Section
5, and the star formation rate density is derived. We discuss these
results and give a summary in Section 6.

Throughout the paper, we use and AB magnitudes and a standard
Cosmology with $\Omega_M=0.3$, $\Omega_{\Lambda}=0.7$ and $h=0.7$.



\section{Search for serendipitous emission lines in the VVDS Deep and Ultra-Deep spectroscopic surveys}

The Vimos-VLT Deep Survey (VVDS) exploited the high multiplex
capabilities of the VIMOS instrument on the ESO-VLT (Le F\`evre et
al., 2003) to collect more than 45000 spectra of galaxies between
$z\sim$0 and $z\sim5$ (Le F\`evre~et~al., 2005;
Garilli~et~al.~2008). In the VVDS-Deep 0216-04 field, more than
$\sim$10000 spectra have been collected for galaxies with $I_{AB}
\leq24$, observed with the LR-Red grism across the wavelength range
$5500<\lambda<9350$\AA, with integration times of 16000 seconds. In
addition, the VVDS Ultra-Deep (Le~F\`evre~et~al.~2010, in preparation)
collected $\sim$1200 spectra for galaxies with $i_{AB} \leq24.75$,
obtained with LR-blue and LR-red grisms, with integration times of
65000 seconds for each grism. This produces spectra with a wavelength
range $3600<\lambda<9350$\AA.  For both the Deep and Ultra-Deep
surveys, the slits have been designed to be 1'' in width, providing a
good sampling of the 1'' typical seeing of Paranal, and between
$\sim$4'' and $\sim$15'' in length, allowing good sky determination on
both sides of the main target. The resulting spectral resolution is
$R\simeq230-250$ for both the LR-blue and LR-red grisms.  The
observations of the VVDS-Deep and Ultra-Deep have been taken in
different observing runs split typically between 3 to 5 nights
respectively. Given the Paranal seeing variations, the seeing quality
of the final observations spans from 0.5'' to 1.2'' FWHM.

In this work, we take advantage of the fact that the VIMOS spectrograph
produces a spectrum for the whole piece of sky covered by each slit.
If a \lae emitting galaxy with a redshift compatible with our wavelength range
serendipitously falls in the slit, the \lae line will appear in the 2-d
spectrum. Obviously other line emitting galaxies at redshifts such that
one or more of their emission line spectrum falls in the observed
wavelength range will also be detected. 

During the processing of the 2-d spectra of the VVDS main targets we
identified that a population of serendipitous emission line objects
was present.  We therefore subsequently performed a systematic search
for serendipitous emission line galaxies in the $\sim8000$ Deep and
$\sim1200$ Ultra-Deep spectra. This was then followed by unambiguous
redshift identification of all the line emitting galaxies as described
in Section \ref{lae:id}.  A significant number of VVDS primary targets
are also \lae emitting galaxies (Le F\`evre~et~al., 2010, in
preparation), we added them to build a more complete LAE sample.

\subsection{The dataset}
\label{dataset}

The dataset consists of three sets of VIMOS pointings: the Deep, the
Ultra-Deep blue, and the Ultra-Deep red. Each pointing
consists of 4 separate quadrants, each containing approximately 100
slits. We summarize each dataset below:
\begin{itemize}
\item Deep dataset: it consists of 20 pointings, for a total of 80
quadrants and about 8000 slits; the main VVDS targets have been
selected from an area of 0.62 deg$^2$ to have 17.5$\leq I_{AB}
\leq$24. The spectra have been obtained with the LR-Red grism,
providing a wavelength range $5500<\lambda<9350$\AA. The exposure times
are about 16000 seconds, allowing to reach fluxes as low as $\sim5
\times 10^{-18}erg/cm^2/s$ (see details below). The total sky
area serendipitously covered is 22 arcmin$^2$.
\item Ultra-Deep dataset: it consists of 3 pointings, for a total of 12
  quadrants and about 1200 slits; the main targets have been selected to have
  23$\leq I_{AB} \leq$24.75 from an area of 0.16 deg$^2$. Each slit
  placed on a primary VVDS target has been observed twice, once with
  the LR-Blue grism, and once with the LR-Red. For the blue spectra,
  the spectral range is $3600<\lambda<6800\AA$, and for the red ones
  it is $5500<\lambda<9350\AA$. Thus, for each of the 1200 slits, the
  Ultra-Deep blue and red spectra overlap for about 1300$\AA$.  We did
  not try to combine spectra in the overlapping regions, because blue
  and red spectra have been observed under different sky background
  and atmospheric seeing conditions. As we will show in the next Sections,
  13 \lae emissions have been found in the overlapping regions. As we
  handled the blue and red spectra separately this allows us to
  compare a posteriori the detection rate, as well as redshifts and
  fluxes.  This is described in the next sections, respectively
  referring to the two samples as 'Ultra-Deep blue' and 'Ultra-Deep red'.
  For each of the blue and red spectra the exposure time is 65000 seconds,
  allowing to reach a flux limit of $\sim1.5x10^{-18}erg/cm^2/s$
  (details below). The total sky area serendipitously covered
  is 3.3 arcmin$^2$.
\end{itemize}

\begin{figure}
  \centering
  \includegraphics[width=8cm]{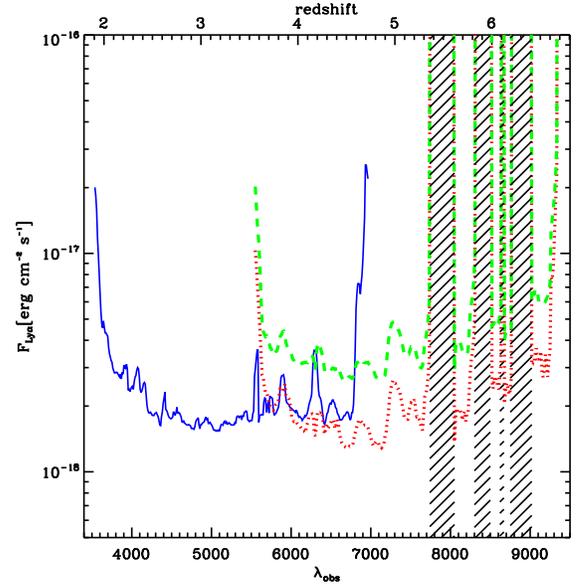}
  \caption{Detection limits for the Ly$\alpha$ line, as a function of wavelength and
      redshift, for the three subsamples: Deep ({\it green dashed line}),
      Ultra-Deep red ({\it red dotted line}) and Ultra-Deep blue ({\it blue
      line}). The shaded regions indicate part of the spectra
      dominated by bright OH sky lines, where the rms of the
      background becomes too high to enable a secure line 
      detection. }
  \label{background}%
\end{figure}

The noise of VIMOS spectra varies as a function of wavelength, as a
result of a combination of different effects.  First, the combination
of the efficiency of VIMOS optics and the CCD quantum efficiency
depends on the wavelength of the incident light. Second, the OH
airglow sky emission lines are in a suite of bands which are
dominating the background for $\lambda>7500\AA$, and are not easy to
subtract from the combined object+sky spectra in low to medium
resolution spectra.  As a result, spectra are noisier at the
wavelengths of the OH bands.  An added uncertainty on the flux level
is also sometimes present at the position of serendipitous \lae
emission because the background subtraction performed by the VIMOS
data processing pipeline VIPGI (Scodeggio~et~al., 2005) is optimized
using the a-priori knowledge of the position of the VVDS primary
target. As the serendipitous target is offset from the main target,
the low order polynomial fit used to remove the background is made
using some background points including the emission line, therefore
artificially lowering the line flux.  Third, the thin E2V detectors on
VIMOS produce fringing which strongly affects the part of the spectra
at $\lambda>$~8000\AA. As the VVDS has been using a low resolution
$R\sim230$ to maximize the number of objects in the survey, this
produces a blend of the OH sky emission features, globally increasing
the background noise at the position of the OH bands. As a result of
these effects, the flux limit varies as a function of wavelength,
hence of redshift.  We determined the theoretical flux limit,
separately for 'Ultra-Deep blue', 'Ultra-Deep red' and Deep, by
measuring the typical rms at different wavelengths in the $2-d$
spectra. Empirically, we decided to set the limit to 3$\sigma$ in
at least 5 contiguous pixels. The limits for the three subsamples are
shown as a function of wavelength in Figure~\ref{background}. It can
be noted that the background, even for each of the three subset, is
changing as a function of the wavelength and redshift. As a result of
the bright OH skylines at $\lambda>$7500\AA, a large part of the
redshift range at $z>5.5$ is in practice inaccessible. However, there
are at least two very clean windows around z=5.7 and z=6.5, with a
very low background corresponding to the absence of OH emission. The
faintest flux limits can be observed in the range $4200 < \lambda <
7600$\AA, reaching down to very faint flux levels of
$\sim1.5x10^{-18}erg/cm^2/s$ at $3\sigma$. The combination of sky area
covered, wavelength range, and depth is unprecedented.

\subsection{Photometry}

The VVDS 0216-04 field benefits from extensive deep photometry.  The
field was first observed with the CFH12K camera in $BVRI$ bands (Le
F\`evre~et~al., 2004; McCracken~et~al., 2003), in the $U$ band
(Radovich~et~al.~2004) and $K$ band (Iovino~et~al.~2005). More recent
and significantly deeper observations have been obtained as part of
the CFHT Legacy Survey in $ugri$ and $z'$ (Goranova~et~al.~2009;
Coupon~et~al.~2009), and as part of the WIRDS survey in $J$, $H$ and
$K$ bands.  The CFHTLS observations reach 5$\sigma$ point source
limiting magnitudes in i band of $i_{AB}\simeq28$ while the WIRDS
observations reach a $3\sigma$ point source limit of $Ks_{AB} \simeq
23.5$ (McCracken~et~al., in preparation).

\section{The \lae sample}

\subsection{Serendipitous emission lines identification}\label{sect:id}

The final database consists of $\sim1200$ 2-d spectra observed with
both red and blue grism (Ultra-Deep dataset), and $\sim8000$ 2-d
spectra observed with the red grism. For each of the 92 quadrants
observed, the reduction code (VIPGI, Scodeggio~et~al.~2005) produces a
frame in which the $\sim$100 2-d spectra are arranged one above the
other, with the spectral direction along the x-axis. With the purpose
of identifying serendipitous lines, we built a data processing
procedure that automatically masked out all the continua and the bad
features (bright residuals and OH skylines). The unmasked area is then
used to search for emission lines, and to compute the effective area
covered by each slit on the sky. An object falling serendipitously in
a slit will produce a spectrum with a combination of continuum and
emission lines. For faint objects with strong lines, the continuum is
not detected, and we are left with emission lines with a line profile
as produced by the spectrograph. For point source objects, VIMOS
produces a line profile along the slit (in the spatial direction)
which is almost identical to the seeing profile (FWHM$\sim1''$ or
about 5 pixels). This is the projected equivalent slit size resulting
from the object size convolved with the seeing profile and the slit
profile on the spectral dispersion direction for a maximum of about 5
pixels corresponding to the 1 arcsecond slit width projection (or
about 30\AA~ spectrally). The 2D line profiles produced are thus
similar to point source images. For compact objects smaller than the
atmospheric seeing disc, and if the seeing is smaller than the 1
arcsec slit width, the observed line profile in the spectral dimension
will be dominated by the projected seeing profile. For extended
objects the line profile becomes wider in the spatial direction
leading to oval shaped emission lines.  As these spectroscopic line
images are similar to those of stars or galaxies on deep images, we
blindly run SExtractor on the masked 2D spectra to automatically
find all emission lines. This was followed by a visual inspection by
two authors (PC and OLF) of all the objects detected by SExtractor,
each producing a catalog of candidates. A further inspection has been
conducted in order to find possible lines missed by SExtractor: we
noted that lines too close to a bright continuum were not found by
SExtractor, probably because of deblending problems.  These two
independent lists were then jointly examined by these two authors to
decide if a candidate was to be kept, with more than 85\% of the lists
being unambiguously and independently selected by both authors. A
fiducial catalog of {\it bona-fide} emission lines has then been built
from this comparison. A more careful determination of the
incompleteness and contamination will be presented in the next
sections. In particular, the completeness is assessed using
simulations in Section~\ref{sect:complsim}, and an estimate of the
contamination is given in Section~\ref{lae:id}.

\begin{figure*}
\centering
\includegraphics[width=15cm]{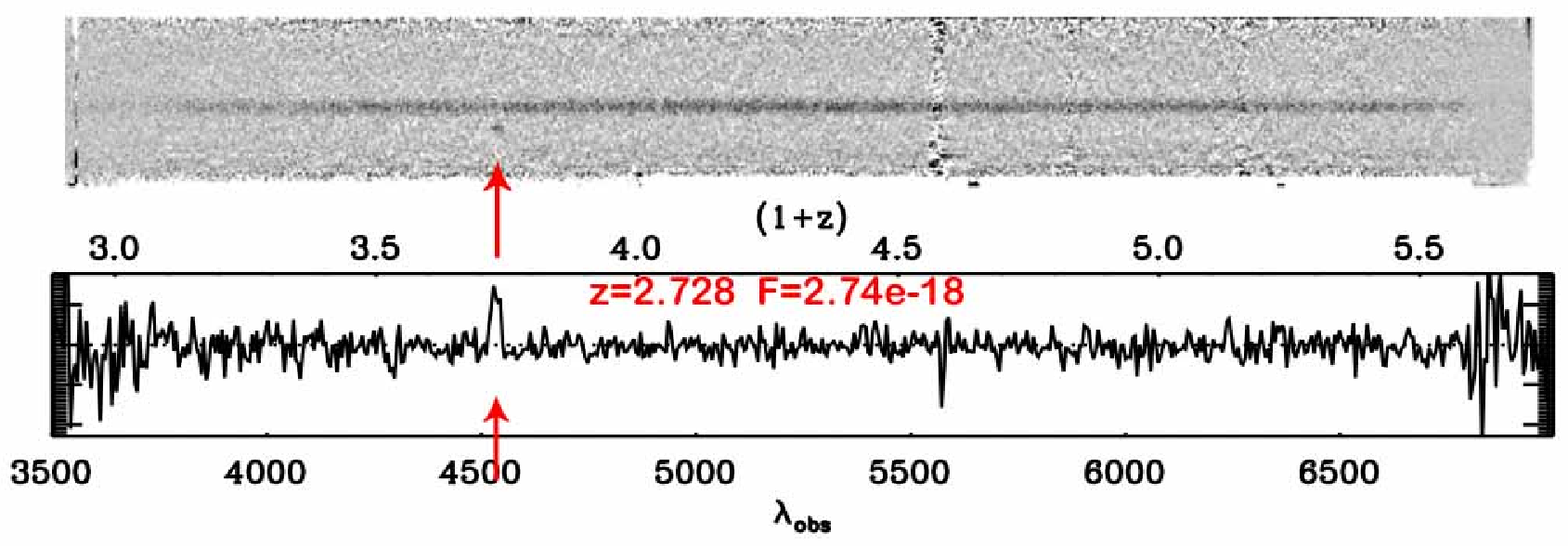}
\includegraphics[width=15cm]{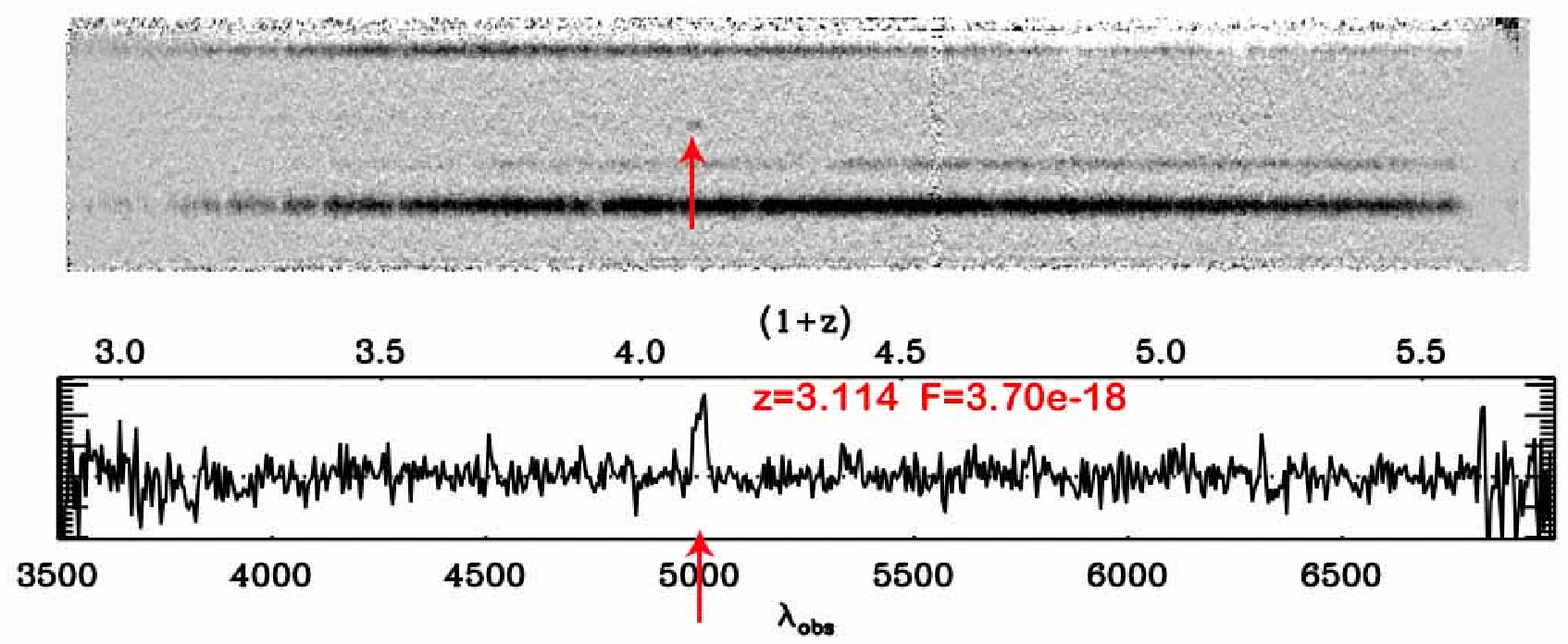}
\includegraphics[width=15cm]{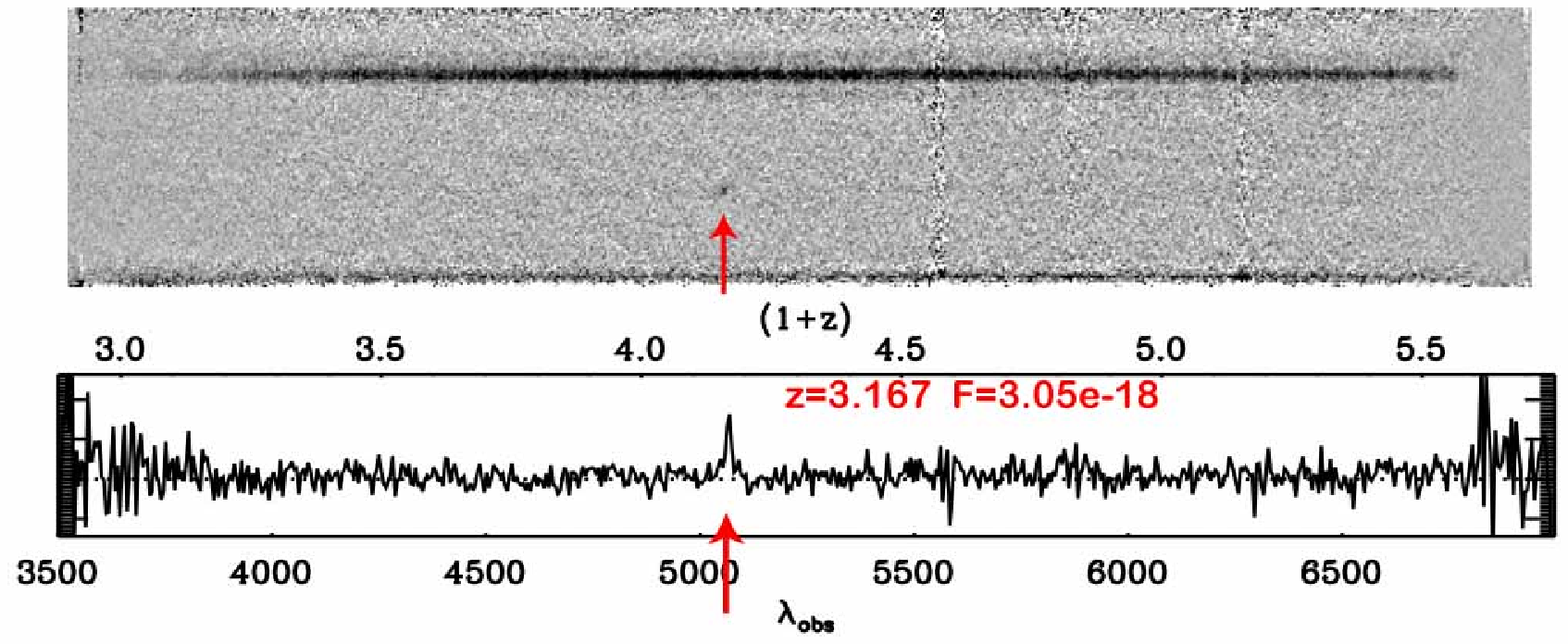}
\includegraphics[width=15cm]{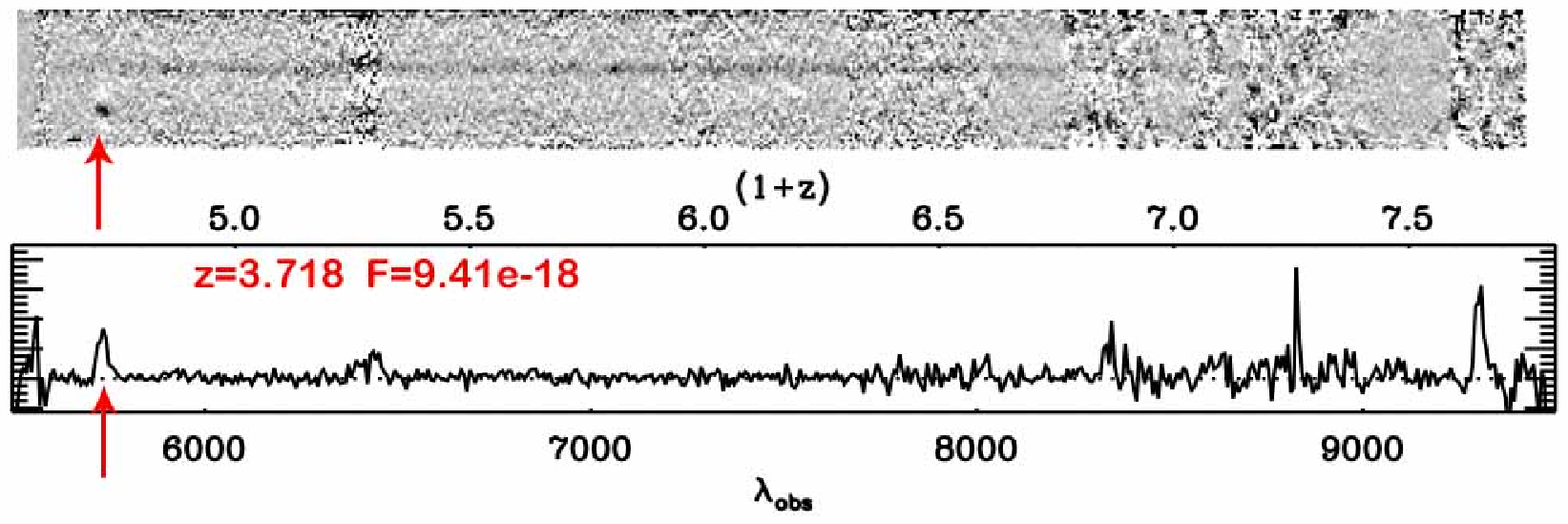}
   \caption{Random selection of LAE candidates from the Ultra-deep
   blue dataset, ordered by increasing redshift. For each of them we show the
   full 2-d (top) and 1-d (bottom) spectra. The position of Ly$\alpha$
   is indicated by an arrow on both the 2-d and 1-d spectra, and by a
   label reporting the redshift on the 1-d spectrum. The observed
   wavelength scale at the bottom of the 1-d spectrum is transformed
   into the redshift of Ly$\alpha$ at each wavelength on the top.}
   \label{spec}
    \end{figure*}

\begin{figure*}
\centering
\includegraphics[width=15cm]{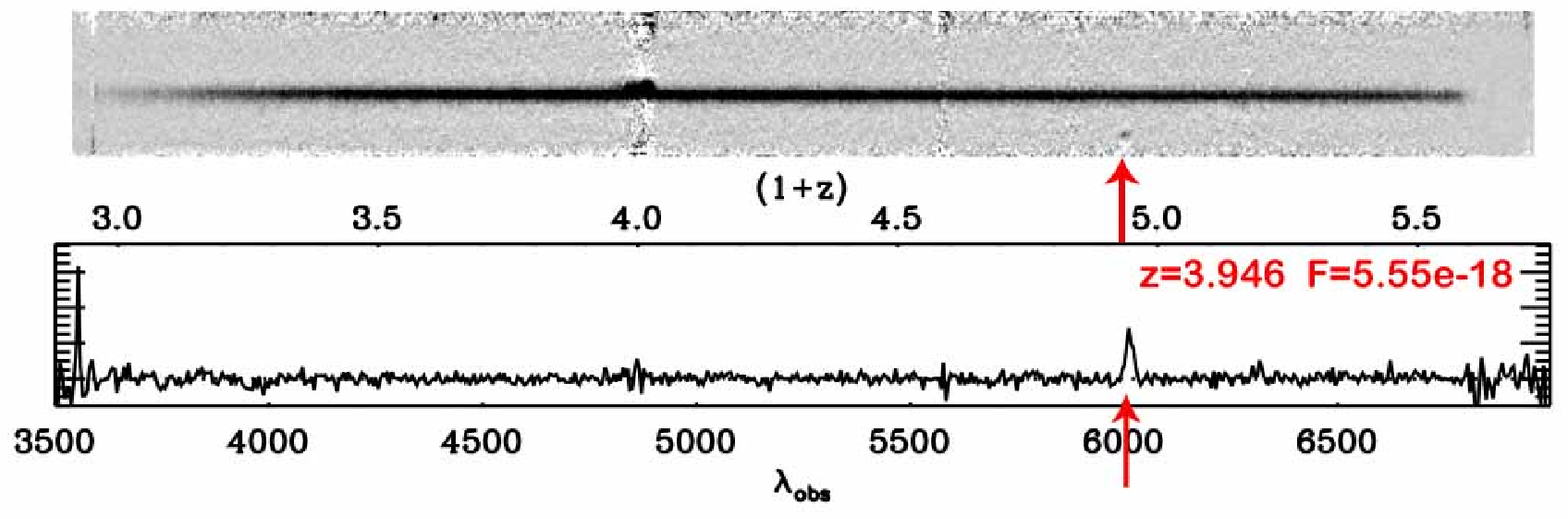}
\includegraphics[width=15cm]{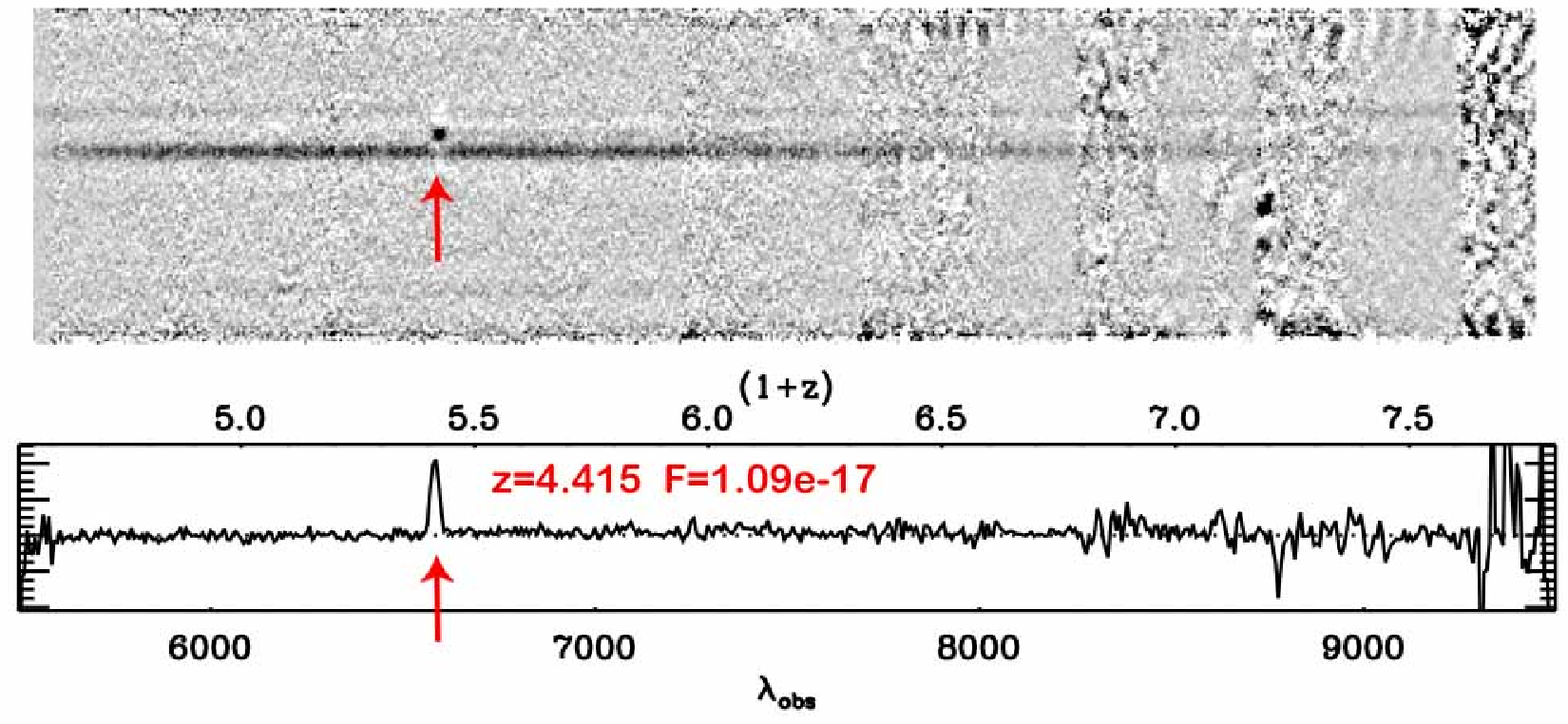}
\includegraphics[width=15cm]{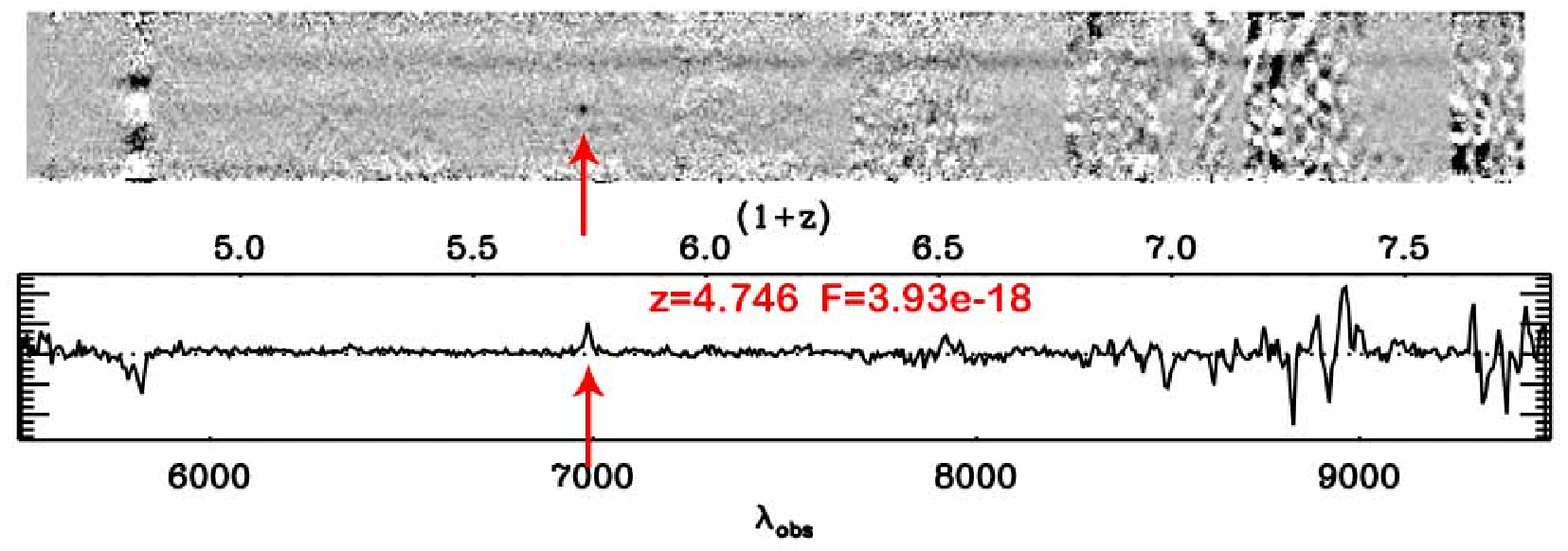}
\includegraphics[width=15cm]{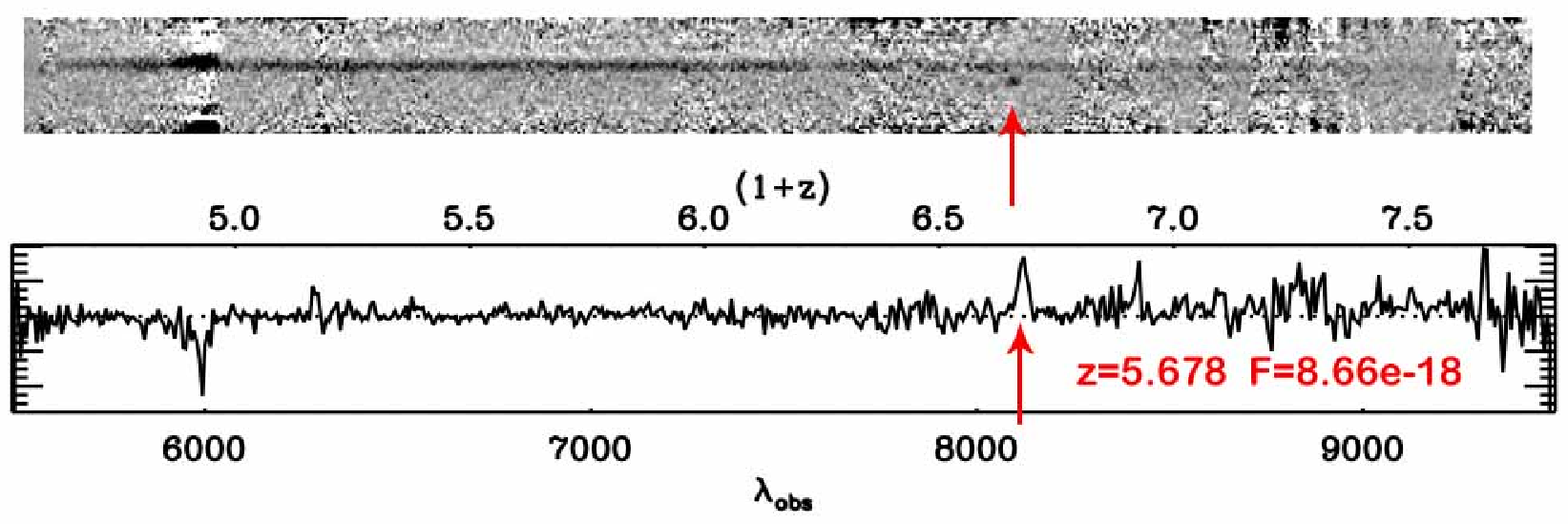}
   \label{spec2}
   \caption{Same as figure~\ref{spec}, for objects detected in the
   Ultra-deep red and in the Deep datasets.}
    \end{figure*}

We obtain a total sample of 133 serendipitous emission
line emitters, 105 found in the Ultra-Deep survey and 28 in
the Deep. Most of these galaxies have no evident continuum appearing
in the spectra. We show a random pick of 8 serendipitous lines in
Figure \ref{spec} and \ref{spec2}. For each line, we show both the 2-d
and 1-d spectra.

\subsection{\lae identification}
\label{lae:id}

Once the fiducial emission line catalog is built, we have to
unambiguously assign a rest-frame wavelength to any given emission
line.  If one emission line is identified in a contiguous spectrum
covering 3600 to 9350 \AA, there are several possibilities for the
line identification assuming a star-forming galaxy spectrum:
below 3727\AA~ the line is most probably \lae, above 3727\AA~ the
emission line can be \lae, [OII]3727\AA, H$\beta$, [OIII]5007\AA~
(assuming [OIII]4959\AA~ is not detected,
[OIII]5007/[OIII]4959$\simeq$3), or H$\alpha$. In the spectrum of
normal star forming galaxies, [OII], H$\beta$ and [OIII] are most
often present. Therefore, as a first approach we assumed that the
emission lines could be [OII] at 3727 \AA, H$\beta$ at 4861 \AA,
[OIII] at 5007 \AA or H$\alpha$ at 6563 \AA, and we checked for
other expected emission lines at different wavelengths in our spectral
window. Using the typical line ratios for star-forming galaxies of
average metallicity we estimate the flux of the other expected lines,
and check whether they could be detected above our flux
limit. However, the line flux ratios strongly depend on the galaxy
metallicity. For example, typical ranges of line ratios are:
[OIII]/[OII]$=0.1\div5$, H$\alpha$/[OII]$=0.3\div5$ and
H$\beta$/H$\alpha=0.1\div0.8$ (Lamareille~et~al.~2006;
Maier~et~al.~2006; Tresse~et~al.~2007; Cowie\&Barger~2008,
Kewley\&Ellison~2008). So, we decided to use the following line
ratios, typical of galaxies with average metallicity:
[OIII]/[OII]$\sim$0.35, H$\alpha$/[OII]$\sim$1 and
H$\beta$/H$\alpha\sim$0.35. Cases of extreme metallicity are discussed
below.

Narrow-line type-II AGN with classical line ratios
would be straightforward to identify given the broad wavelength
range of our spectra. If we were to assign the single emissions lines 
to lines like CIV-1549, CIII-1909, or MgII-2800 we would expect to 
see other emission lines in our  observed range either towards the
UV (Ly$\alpha$, CIV, CIII) or towards the red (MgII, [OII], [NeIII]). 
Broad line type-I AGN are also relatively easy to identify
at our spectral resolution.  We have not found any type-I or type-II AGN 
in our data, consistent with expected AGN counts in the area sampled
(e.g. Bongiorno~et~al., 2007; Lamareille~et~al., 2009).

As the spectral window spans from 3600\AA~ to 9350\AA~ for the
Ultra-Deep subsample and from 5500\AA~ to 9350\AA~ for the deep
subsample, we analyze these samples separately.

\begin{itemize}
\item Ultra-Deep: 2 single lines in the Ultra-Deep are at
$\lambda<$3727\AA, and then can likely be identified as \lae, while
103 are at $\lambda>$3727\AA.  Among the latter, 93 lines are at
$\lambda<$6920\AA. If they were [OII], we would to have [OIII] and
H$\beta$ in our spectral window. We estimate, using the line ratios
reported above, that 75 of these 93 lines should be accompanied by
either detectable [OIII] or detectable H$\beta$ or both, but as they
are not detected they are therefore most probably to be identified
with \lae. For the 10 emission lines at $\lambda>$6920\AA, this
argument cannot be used, because the possible [OIII] and H$\alpha$
emission would lie out from our spectral window. Using similar
arguments, we can infer that none of the Ultra-Deep lines are
H$\beta$, [OIII] or H$\alpha$, because they would be always
accompanied by at least one other line in our spectral window with a
flux brighter than the flux limit.
\item Deep: The deep subsample contains 28 serendipitous lines. Again,
all of them can be in principle interpreted as both \lae or [OII]. Since
for the Deep the spectral window is smaller than for the Ultra-Deep,
the same argument as for the ultra-deep tells us that only 7 of these
lines are not [OII]. For the other 21, either a possible [OIII] emission
is outside the spectral range, or it would be too faint to be
detected. However, we can conclude also that the bulk of them are not
H$\alpha$, [OIII] or H$\beta$, because at least one other detectable
line would be detected in our spectra.
\end{itemize}

From this first analysis, we can conclude that basically none of the
serendipitous lines in the sample are H$\alpha$, H$\beta$ or [OIII],
that 84 among the 133 {\it bona fide} serendipitous lines are not
[OII], and that most of the line identification uncertainty for the
other 49 emission lines is between [OII] and \lae.  We note that these
findings will not change too much if we had used much lower values for
the metallicity. Assuming, for example, [OIII]/[OII]$\sim$10, if the
serendipitous line was assigned to [OII], we should always have detected
[OIII] for all lines with $\lambda<6920$.

\begin{figure}
\centering
\includegraphics[width=1.0\linewidth]{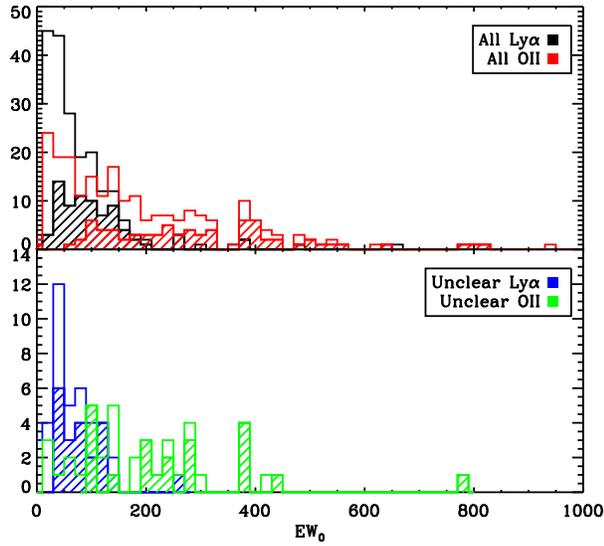}
\caption{ {\it top panel}: For the full sample, the distribution of
rest-frame EW if the lines are identified as Ly$\alpha$ (black
histogram) or as [OII]3727 (red histogram). The filled histograms show
objects without a detected continuum, for which the EW is a lower
limit.  {\it lower panel}: the same as the top panel, but for the 49
lines identified as ambiguous. The blue and green histograms represent
respectively the rest-frame EW if the lines are identified as
Ly$\alpha$ or as [OII]3727.
}
\label{hist_ew}
\end{figure}

We can perform a further check in order to evaluate the contribution
of \lae and [OII] emission to the observed population.  In particular,
we measured the observed equivalent width of the lines to produce
the distribution of the rest-frame EW for each of the two cases, as
presented in Figure \ref{hist_ew}, for all the 217 lines as well as
for the 49 ambiguous lines. While the distribution of the rest-frame EW for
the lines if they are Ly$\alpha$ is mostly below 150\AA, we find that
if the lines were assigned to [OII], $\sim$70\% of the distribution
would be with EW[OII]$>100$\AA.  Among the 49 galaxies that cannot be
unambiguously classified as either \lae or [OII], 37 would have
EW[OII]$>100$\AA. We compare this distribution to the distribution of
EW[OII] observed for galaxies with $M_B \simeq -18$ in the VVDS-Deep
(Vergani~et~al.~2008) in the same redshift range $0.86 < z < 1.5$, and
we note that 90\% of the normal galaxy population has
EW[OII]$<100$\AA. At $z\simeq1$, galaxies with EW[OII]$>100$\AA~ have
only rarely been observed in deep spectroscopy surveys, and none with
EW$>150$\AA~ (Hammer~et~al., 1997; Vergani~et~al., 2008). Since we
expect to have just 5 lines with EW[OII]$>100$\AA~ (10\% of 49), we can
conclude that for about 32 (37-5) the identification is very likely to
be \lae. So, at the most we can identify the line as [OII] for 17
galaxies (49-32). We therefore conclude that about 10\% of the single
emission lines could be assigned to either [OII] or Ly$\alpha$, 
while more 90\% of them are most likely Ly$\alpha$.

\begin{small}
\begin{figure}
\centering
\includegraphics[width=1.0\linewidth]{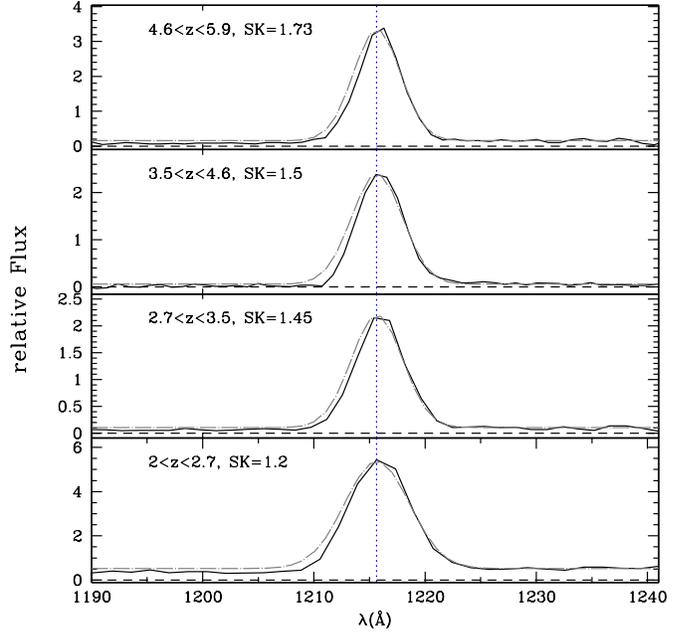}
\caption{Combined spectra for all serendipitous LAE in 4 redshift
domains, centered on Ly$\alpha$. The asymmetry of the line profile is
clearly seen in each redshift range when comparing to a gaussian fit
of the red wing of the line (long-dashed line). For each combined
spectrum we report also the skewness calculated in the range $1201 <
\lambda < 1231$\AA.}
\label{all_lya_z}
\end{figure}
\end{small}

\begin{small}
\begin{figure}
\centering
\includegraphics[width=1.0\linewidth]{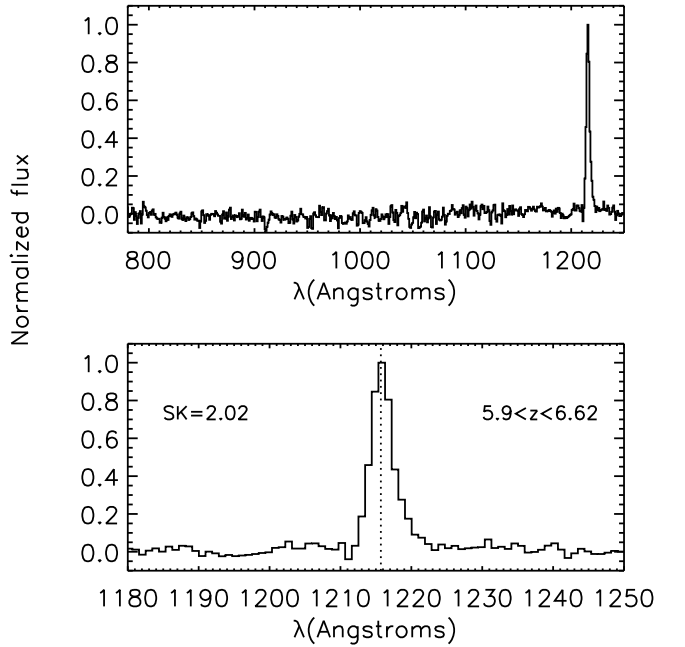}
\caption{Combined spectra for $5.9<z<6.62$. The top panel shows the
continuum bluewards of the \lae line. The bottom one focuses on the
line, showing the asymmetric profile. The value for the skewness of
the line is also reported.  }
\label{all_z6_lya}
\end{figure}
\end{small}

Another check on the line identification with \lae is to look at the
line profile. Ly$\alpha$ line profiles are most often asymmetric, with a
truncation towards the blue because of absortion by the intergalactic
medium, and an extended red wing because of complex velocity fields
(Shapley~et~al., 2003; Ouchi~et~al., 2008). On individual objects, the
line detection is not always of sufficient S/N to detect a line
asymmetry, instead we produced the average normalized \lae line
profiles for LAE in different redshift ranges up to $z\sim6$, as shown
in Figure~\ref{all_lya_z} (the description of the combination
technique and a more detailed analysis of the combined spectra are
presented in Section~\ref{sect:spectropro}). The asymmetry of the
profile is clearly seen at all redshifts, even in the lower redshift
bins. For each combined spectrum we measured also the skewness of the
line in the range $1201<\lambda<1231 \AA~$, that measures the degree
of asymmetry of a given distribution. If the line profile would be
perfectly described by a gaussian, that is by definition symmetric,
the skewness would be zero. On the other hand, a positive value for
the skewness indicates a distribution skewed towards red
wavelengths. As Figure~\ref{all_lya_z} shows, in each redshift bin
between $2<z<5.9$ the skewness is positive and is increasing with
redshift as expected from IGM absorption, reinforcing the primary
identification with \lae.

There are 6 galaxies with a single emission line in the atmospheric
windows $8450 < \lambda < 8620$ and $9000 < \lambda < 9300$ clean of
OH sky emission lines, and they observed equivalent widths $540 <
EW < 1250$\AA.  If these lines were [OII]3727 or H$\alpha$, the rest
EW would be $220 < EW < 510$ or $400<EW<880$\AA~ with a mean EW=315 or
550\AA~ respectively, clearly outside the range of EW for [OII] or
H$\alpha$ emitting galaxies at $z\sim1$. No sign of the broad
component of an AGN which could produce high H$\alpha$ EW has been
observed.  We also produced the line profile of the 6 galaxies in this
redshift range, reported in Figure~\ref{all_z6_lya}. The resulting
profile is clearly asymetric, as confirmed by the positive value for
the skewness. As other single emission line candidates with strong
observed EW are even more unlikely than [OII]3727 or H$\alpha$ and
should show other emission lines in our observed wavelength domain, we
argue that the most likely possibility is that these emission lines
are Ly$\alpha$ with $5.96 < z < 6.62$, making them some of the most
distant galaxies found to date.

In summary, among the 133 emission lines identified serendipitously in
the 2D spectra, we conclude that 124 are most likely \lae, while a
low fraction of 14 (10\%) can be either \lae or [OII]. In the
following, we assume that all galaxies with ambiguous line assignment
are \lae.

\subsection{Final sample}
We completed the LAE sample by adding the primary VVDS
spectroscopic targets that unambiguously have \lae in emission. We
found 70 \lae emitters among the $\sim1200$ Ultra-Deep primary
targets, all at redshift $z<3.5$, and 14 among the $\sim$8000 Deep
targets. The Deep survey is actually limited to $\lambda>$5500\AA,
hence it could not see \lae at $z<3.5$. All these galaxies have a high
quality flag (larger than or equal to 2; Le F\`evre~et~al., 2010, in
prep.) for the redshift, since they have other spectral features that
allow to unambiguously identify the redshift and therefore classify
them as \lae emitters. They show in fact OI at 1303\AA~, CII at
1333\AA~, SiIV at 1397\AA~, CIV at 1549\AA, and sometimes CIII at
1909\AA~. The final sample is therefore made of 217 LAE, including
133 serendipitous LAE, and 84 LAE from the primary VVDS targets. The
sample is summarized in Table~\ref{tab:sample}.

\begin{table}
 \centering
\begin{tabular}
{clr@{$\pm$}rr@{$\pm$}lr@{$\pm$}lr@{$\pm$}lr@{$\pm$}lc}

\hline\hline \noalign{\smallskip} 


\multicolumn{1}{c}{} & \multicolumn{2}{c}{U-DEEP} & \multicolumn{2}{c}{U-DEEP} & \multicolumn{2}{c}{DEEP} & \multicolumn{2}{c}{total}\\ 
\multicolumn{1}{c}{} & \multicolumn{2}{c}{blue} & \multicolumn{2}{c}{red} & \multicolumn{2}{c}{} & \multicolumn{2}{c}{}\\ 

\noalign{\smallskip} \hline
\noalign{\smallskip}

\multicolumn{1}{c} {Serend.}& \multicolumn{2}{c}{94$^{1}$(46)} & \multicolumn{2}{c}{24$^{1}$(7)} & \multicolumn{2}{c}{28(13)} & \multicolumn{2}{c}{133(66)}\\ 
\noalign{\smallskip}
\multicolumn{1}{c} {Targets}& \multicolumn{2}{c}{70} & \multicolumn{2}{c}{0} & \multicolumn{2}{c}{14} & \multicolumn{2}{c}{84}\\ 

\noalign{\smallskip} \hline
\noalign{\smallskip}
\end{tabular}
\caption{The final sample of \lae emitters, divided into serendipitous
 objects and spectroscopic targets. The number in parenthesis for the
 serendipitous galaxies indicates the objects with optical
 counterpart. Note:$^{1}$ 13 lines have been observed in both U-DEEP
 blue and U-SEEP red}\label{tab:sample}
\end{table}

For each emission line, we carefully measured the position of the line
(and thus the redshift), the total flux in the line and the
continuum. The continuum, in units of $F_{\lambda}$, is measured
as close as possible to the red wing of the \lae line, by averaging
the counts between 1230 and 1250\AA. From these fundamental
quantities, the observed and rest-frame equivalent widths and the
luminosities are derived:
\begin{equation}
L=4\pi F d_l(z)^2\\
EW_{rest}=F/C\frac{1}{1+z}
\end{equation}
where $L$ and $F$ are respectively the luminosity and the flux in the
line, $z$ is the redshift of the line, $d_l(z)$ is the luminosity
distance at the redshift of the line, $EW_{rest}$ is the rest-frame
Equivalent Width and C is the continuum around the line.  Obviously,
when the continuum is not detected, the derived equivalent widths are
just a lower limit, and we used 1 sigma of the background as the
continuum value.

Moreover, each luminosity can be converted to a star formation rate
(SFR) combining the classical prescription to derive the SFR from
H$\alpha$ Luminosity by Kennicutt~(1998) and the expected ratio
between Ly$\alpha$ and H$\alpha$ flux. In particular:
\begin{equation}
SFR(M_{\odot}yr^{-1})=L_{H\alpha}(ergs~s^{-1})/(1.26\times10^{41}),
\end{equation}
with H$\alpha$ flux corrected for internal extinction and valid for a
Salpeter IMF. The conversion between H$\alpha$ and Ly$\alpha$
luminosity has been theoretically derived by Brocklehurst~1971
(assuming case B recombination):
\begin{equation}
L_{Ly\alpha}=8.7L_{H\alpha},
\end{equation}
and it does not account for dust and escape fraction corrections.

Combining the two gives:
\begin{equation}\label{eq:sfr}
SFR(M_{\odot}yr^{-1})=L_{Ly\alpha}(ergs~s^{-1})/(1.1\times10^{42}).
\end{equation}
   \begin{figure}
   \centering\includegraphics[width=6cm]{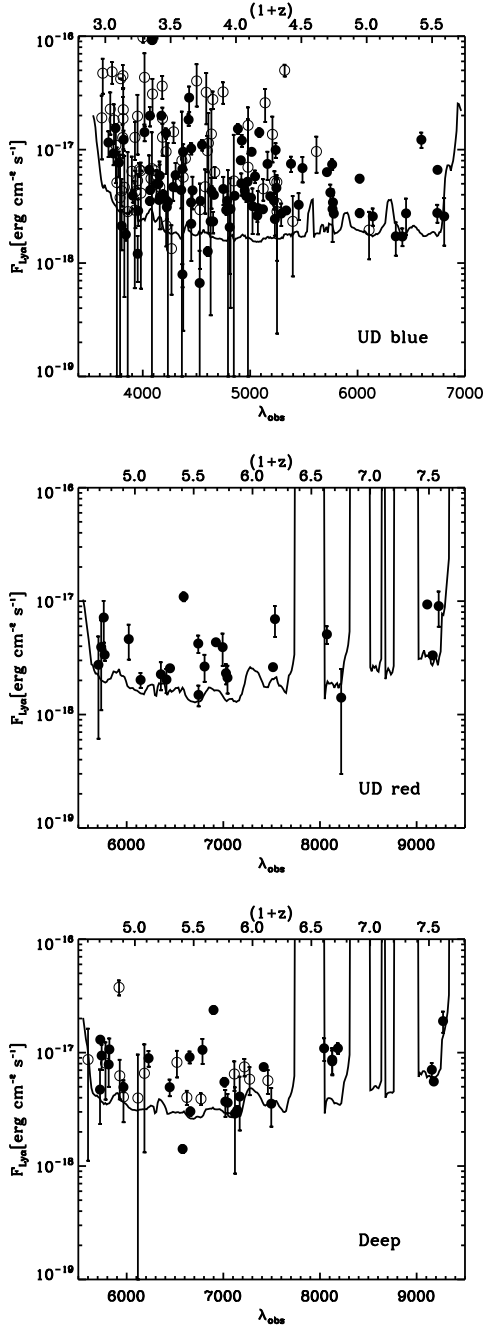}
   \caption{For each sub-dataset, we show the flux versus redshift
      together with the flux limit as a function of the redshift and
      wavelength ({\it solid lines}). The flux limit is
      empirically measured on the 2-d spectra. Filled and open
      circles represent respectively serendipitous \lae galaxies and
      targets with \lae emission.  }
   \label{flux_limits}%
    \end{figure}

In Figure \ref{flux_limits} we report the flux and the position (both
wavelength and redshift if the lines are identified as \lae) for the
three surveys separately. It can be noted that the Ultra-Deep blue is
providing the biggest sample with 164 emission lines galaxies (94
serendipitous lines and 70 targets of the spectroscopy), while the
Ultra-Deep red and the Deep contribute with respectively 24 and 42
emission lines galaxies. Note also that in a few instances we measure
line fluxes that are weaker than the nominal flux limit at that
wavelength: this is because the background, at each wavelength, is
measured as the average of the backgrounds in the different quadrants
and pointings. For some pointings, the background is lower than the
average, and thus fainter lines can be detected.

\subsection{Completeness simulations}\label{sect:complsim}

As we saw in Fig.~\ref{background}, the background of our spectra has
a complicated structure, as there are strong variations with
wavelength.  To perform a statistical analysis of the number density
evolution of \lae galaxies, it is important to carefully determine,
for each wavelength (and thus for each redshift), what is our
capability to recover a line with a given luminosity.

We performed a Monte Carlo simulation, building a catalog of 1000
fake lines for each dataset, for a total of 3000. Each line has a
known flux, randomly extracted from a flat distribution between
$F_{Ly\alpha})=1\times10^{-18}$ and $F_{Ly\alpha})=1\times10^{-16}$. The
redshift is randomly extracted from a flat distribution between z=2
and and z=4.7 (for blue spectra) and between z=3.7 and z=6.7 (for red
spectra).  The fake lines are then added at the wavelength
corresponding to the known redshift, in a randomly extracted
slit. Since emission lines in our sample are not or barely resolved
either in the spatial or in the spectral directions (see next
sections), we chose to use simple gaussian profiles with FWHM of
1''$\times$30\AA.

   \begin{figure}
   \centering
   \includegraphics[width=8cm]{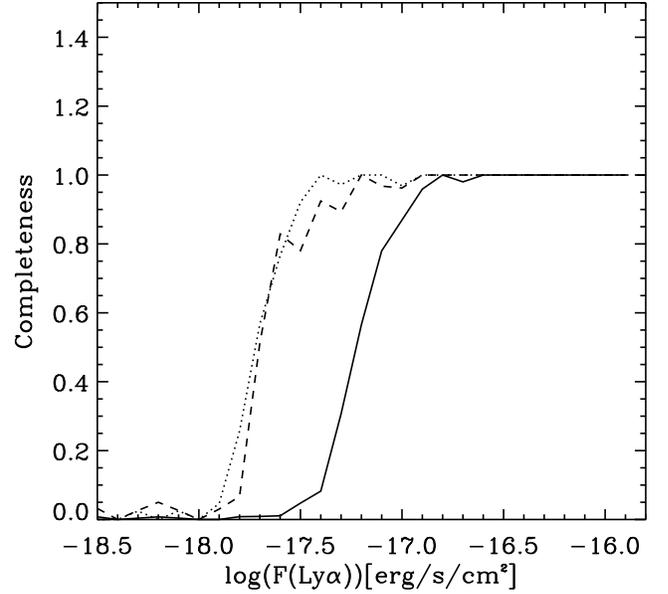}
   \caption{
Completeness in line identification from Monte Carlo
simulation as a function of the flux, for the three subsamples:
Deep ({\it solid line}), Ultra-Deep red ({\it dotted line}) and
Ultra-Deep blue ({\it dashed line}).}
   \label{completeness}%
    \end{figure}

We attempted to recover these simulated lines, without a-priori
knowledge of redshifts and fluxes, by applying the same procedure used
on the real 2-d spectra, described in Section~\ref{sect:id}. The
results are reported in Figure \ref{completeness} presenting the
completeness as a function of the line luminosity and of the redshift,
for the three subsamples. We can see that Ultra-Deep blue \& red and
Deep are 50\% complete respectively at $F_{Ly\alpha}\sim
2\times10^{-18}$ and $\sim 8\times10^{-18}$.


\subsection{Slit flux loss}\label{sect:loss}

Since the slits used have sizes comparable to the seeing of the
images, part of the line flux cannot be recovered and will be always
lost. The flux loss depends on the offset between the galaxy and the
slit center positions, on the size of the galaxy with respect to the
slit width and on the atmospheric seeing.  For an object perfectly
centered in the slit, the flux loss will be minimal, but not zero. If
the position of the object in the slit is known, as in the case of the
objects with an optical counterpart, it is relatively easy to compute
the flux loss. However, a significant number of galaxies in our sample
do not have an optical counterpart within $\pm$0.5'' from the expected
position of the object along the slit and within $\pm$2'' off axis in
the direction perpendicular to the slit (see Section
\ref{section:counterparts}).

As we do not know for these galaxies their positions inside the slits,
we cannot compute object by object the actual flux loss. This problem
is common to all surveys of serendipitous \lae emitters
(e.g. Rauch~et~al.~2008; Lemaux~et~al.~2009), and is usually solved
with a statistical approach for the objects with no detected
counterpart.  In this Section, we derive the slit flux loss as a
function of the position of the object in the slit, so that we can
correct the line flux for those objects with an optical counterpart,
and then we estimate a statistical correction for the objects without
any optical counterpart.

\begin{figure}
   \centering
   \includegraphics[width=8cm]{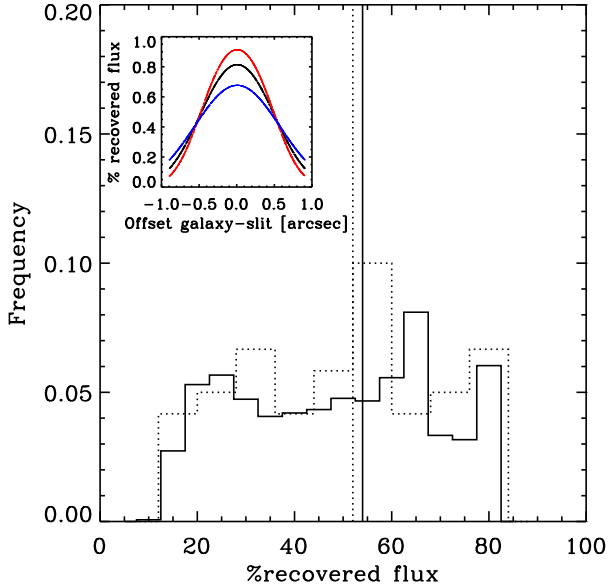}
   \caption{Histogram of the percentage of recovered flux, for a set
of a 3000 simulated lines randomly offset with respect to the center
of the slit (solid line) and for the serendipitous lines (dotted
line). The vertical lines show the median for the two distributions,
around 55\% of the recovered flux. The small inset shows the recovered
flux as a function of the offsets for the simulated lines, for the
0.7, 0.9 and 1.2 arcsec FWHM lines (respectively in red, black and
blue).}
   \label{slit_loss}%
    \end{figure}

We start from the consideration that the bulk of our \lae galaxies are
unresolved in the spatial direction (see
Section~\ref{section:counterparts}), having sizes comparable with the
seeing of the observations. The FWHM distribution in the spatial
direction of the lines in our sample spans from 0.5 to 1.2 arcseconds,
with 90\% of them having FWHM=0.7$\div$1.2 arcsec (see
Figure~\ref{size_dist}). This implies that we can model them as
gaussians, with fixed FWHM. We built a simulation in which an emission
line object with a given FWHM is placed in a 1 arcsec slit at
different positions with respect to the center of the slit until it is
completely outside of the slit width. The small inset in
Figure~\ref{slit_loss} shows the fraction of recovered flux as a
function of the offset between the objects and the slit, for three
seeing conditions FWHM= 0.7, 0.9 and 1.2 arcsec (corresponding to
$\sigma\sim$0.3, 0.38 and 0.5 arcsec). If the galaxy was perfectly
centered in the slit, only $\sim70$, 80 and 90\% of the flux would be
recovered, respectively for the 1.2, 0.9 and 0.7 arcsec FWHM lines; on
the other hand, if the galaxy would be placed at 2$\sigma$=0.8 arcsec
outside the slit, only 20\% of the light would be retrieved.

Thus, we can estimate that the slit flux loss for the 84 primary
spectroscopic targets, that are perfectly centered in the slits, is
around 15\%, with a typical range between 10\% and 30\% depending on
the seeing. For the serendipitous object with an optical counterpart,
whose offset with respect to the center of the slit is known, we
estimated the flux loss assuming they are compact under the average
seeing conditions of a pointing. We show their distribution of the
recovered flux in Figure~\ref{slit_loss}.

In order to check if the percentage of recovered flux that we find for
the serendipitous sample is reasonable, we use a Monte Carlo
simulation. In particular, we generate 3000 2-dimensional gaussian
profiles with FWHM randomly extracted between 0.7 and 1 arcsec,
applying an offset between the position of the peak and the center of
the slit randomly extracted between $\pm$1 arcsec.  The distribution
of the recovered flux for the 3000 lines is shown in
figure~\ref{slit_loss}. We note that the choice of $\pm1$'' as maximum
offset introduces an artificial cut-off at $\sim$10\% of the recovered
flux. This choice however is justified by the fact that an
hypothetical object placed offslit by more than $\pm1$'' should have a
true \lae flux $>10\times$ stronger than the one measured in the slit
spectroscopy, hence its optical counterpart should be visible in
photometry. There is an overall agreement from the result of this
simulation and the flux loss estimated for the serendipitous lines
with optical counterpart. For $\sim$40\% of the cases, the recovered
flux is higher than 70\%. The median of both histograms is
$\simeq55$\%: we will use this value to correct the measured fluxes
for the serendipitous lines with no optical detection in our sample.

\section{Sample properties}
\subsection{Redshift distribution}
\begin{figure}
   \centering\includegraphics[width=8cm]{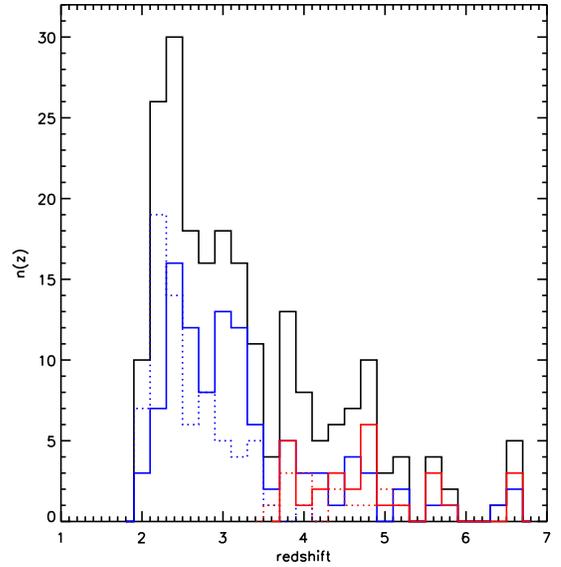}
   \caption{Redshift distribution for the whole sample (black line),
     as well as for the different subsamples: solid lines show
     serendipitous lines, while dotted ones indicate primary
     spectroscopic targets; blue and red lines show the Ultra-Deep and
     deep subsamples, respectively.}
   \label{dz}%
    \end{figure}

In Figure \ref{dz} we show the redshift distribution for our
sample. The distribution extends from $z=2$ to $z=6.7$, with a median
$<z>$=3. We have 5 detections at $z\sim6.5$ and 9 at $z\sim5.7$. The
main VVDS targets with \lae come mainly from the Ultra-Deep survey,
and interestingly, they show a different redshift distribution: even
though the peak for the two subsample is around $z=2.2\div2.5$, the
serendipitous show a broader tail at $z>$4, while 85\% of the targets
have $z<4$. A simple Kolmogorov-Smirnov test rules out with more than
99\% confidence that the two distribution are statistically
equivalent. This reflects the different selection for the 2
subsamples: magnitude selected for the former, flux selected for the
latter.

\subsection{Optical counterparts and size distribution}\label{section:counterparts}

   \begin{figure}
   \centering\includegraphics[width=6cm]{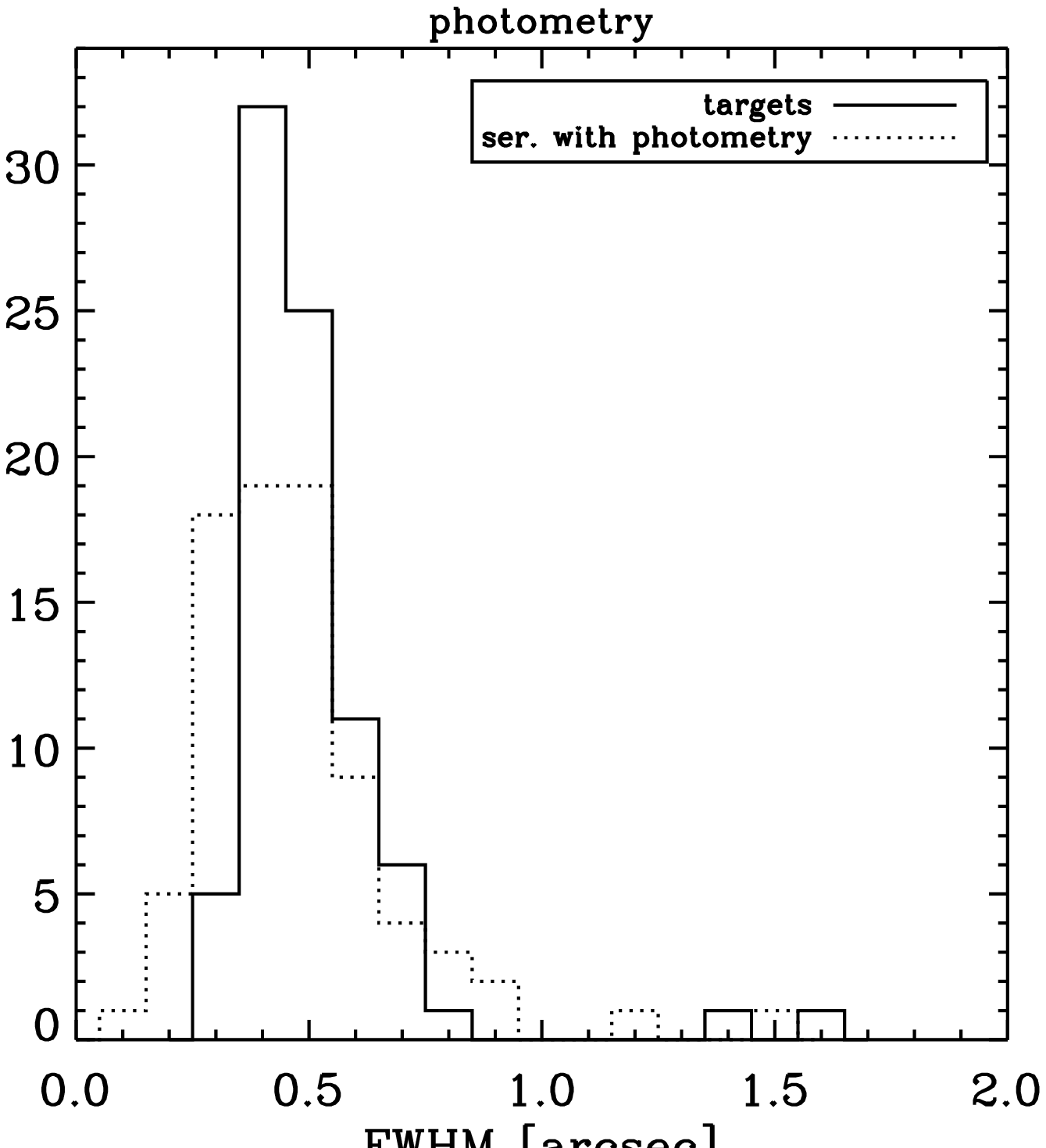}
   \centering\includegraphics[width=6cm]{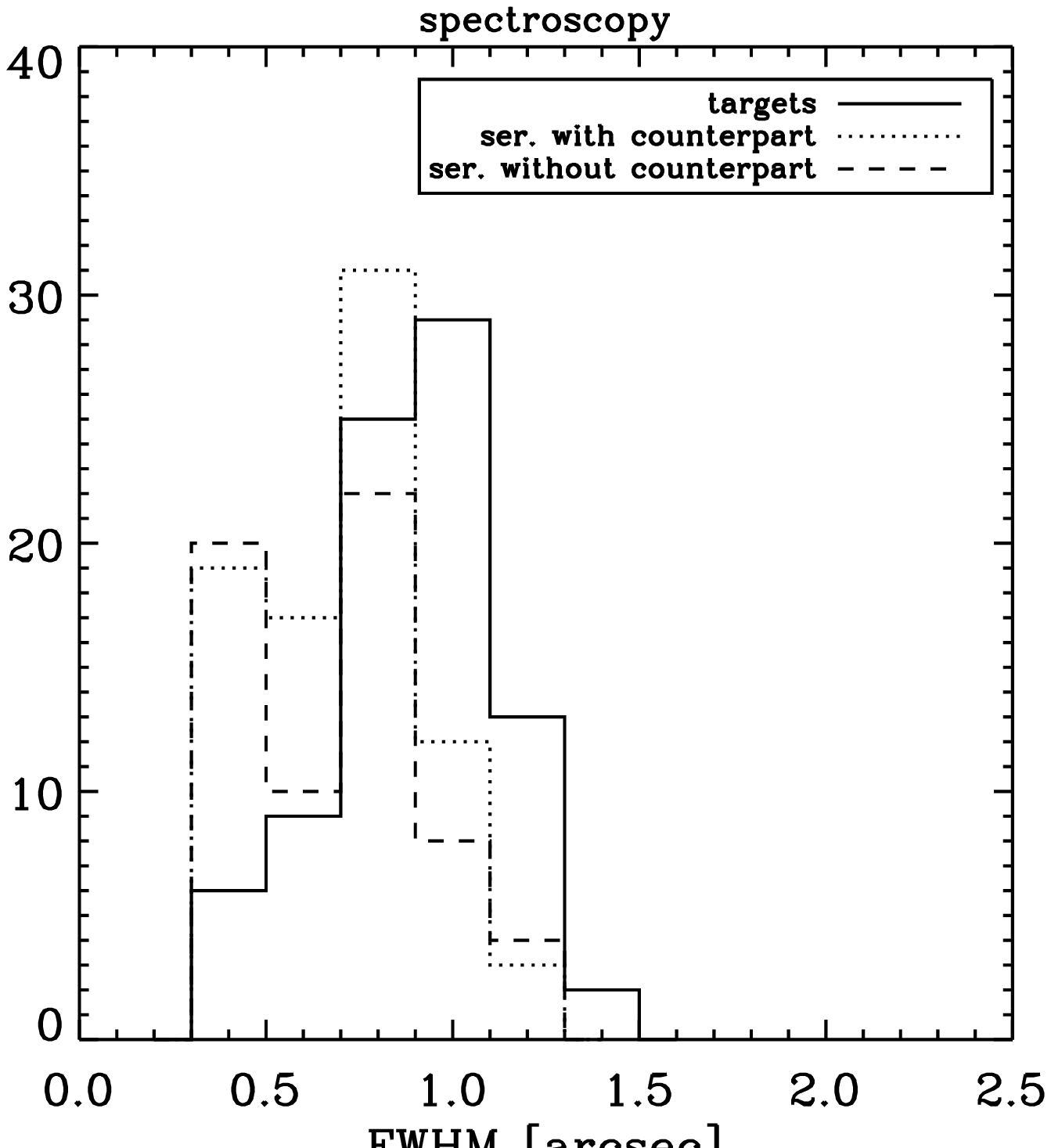}
   \caption{Full width half maximum (FWHM) distribution as measured
from the CFHTLS images (top panel) and from the Ly$\alpha$ 2-d spectra
(bottom panel). Targets are shown with a solid line,
serendipitous objects with and without photometric counterparts are
shown with a dotted and dashed line, respectively. The CFHTLS image
used has being built by stacking the 25\% best-seing exposures, and
has a final PSF FWHM of about 0.5''. In the top panel, we just show
serendipitous objects with photometric counterpart.}
   \label{size_dist}%
    \end{figure}

We searched the deep CFHTLS ugriz images, including the combined ugriz
stack, for optical counterparts of the 133 serendipitous LAE. In order
to account for possible positional uncertainties, we required the
counterpart to be within $\pm0.5''$ from the position of the spectrum
along the slit. Moreover, since even bright off-slit objects can show
a detectable spectrum (see Sect.~ \ref{sect:loss}), we search for
counterparts within $\pm2''$ of axis. We found faint counterparts for
53 (50\%) and 13 (46\%) LAE respectively in the ultradeep and in the
deep sample, with magnitudes $i_{AB}$ ranging from 23.5 to 27.5.
These counterparts are always $\pm$1'' off axis at the most. For the
remaining 67 objects we did not find an optical counterpart down to a
magnitude $AB\sim28$.

In Figure~\ref{size_dist} we compare the Full Width at Half Maximum
(FWHM) for the serendipitous \lae emitters and the spectroscopic
targets, as measured on the deep CFHTLS images and directly on the 2-d
spectra of the Ly$\alpha$ line.

In particular, we used CFHTLS-D1 (VVDS-02h field) D-25 stack in the
$z-$band, that has been built stacking the 25\% best seeing images
together resulting in a PSF FWHM of about 0.5 arcsec (see Table 26 of
Goranova~et~al.~2009). Thanks to the better angular resolution, these
images are very useful to check whether or not our galaxies are
resolved in the spatial direction. On the other side, we are also
interested in measuring the size for our sample also on the 2-d
spectra, to check the assumption we made in
Sect.~\ref{sect:loss}, where we estimated the slit flux loss.

The FWHM of the serendipitous and main VVDS magnitude selected
populations with optical counterparts measured in the D-25 stack are
compared in the top panel of Figure~\ref{size_dist}. The distribution
for both populations peaks at around 0.5 arcsec with a gaussian
distribution width. This is expected from the measurement errors on
these faint objects if they are unresolved under the seeing of the
D-25 stack. The width of the Ly$\alpha$ line measured on the 2-d
spectra (bottom panel of Figure~\ref{size_dist}) indicates that the
Ly$\alpha$ line is emitted from a compact region, as the size
distribution is comparable to the seeing distribution of the
spectroscopic observations. This is therefore indicating that most
of the LAE are compact, both on the continuum and line emission.

From these measurements it is clear that the faint LAE population is
compact in size.  Interestingly, there is a clear lack of large
objects (FWHM$>$1--1.5 arcsec) in our sample. Faint LAE clearly have
different sizes than the large \lae blobs identified in the bright LAE
population (Steidel~et~al.~2000; Matsuda~et~al.~2004; Ouchi~et~al.,
2008).  The weak continuum and Ly$\alpha$ sizes are in agreement with
Venemans~et~al. (2005), Taniguchi~et~al. (2009), and the HST-based
study of Bond~et~al. (2010).  Our results support a picture where
Ly$\alpha$ emission from faint LAE is compact and originates from the
same area as the UV continuum as found by Bond~et~al. (2010), and we do
not confirm the trend reported by Nilsson~et~al. (2009) of more
extended sizes in the narrow band Ly$\alpha$ images than in the
broad-band images. While extended Ly$\alpha$ emission like observed in
Ly$\alpha$ blobs and Ly$\alpha$ halos is expected if Ly$\alpha$ is
produced from resonant scattering on diffuse gas surrounding the
galaxy, our data may indicate that the gas reservoir of faint LAEs is
rather compact or that, if extended, it is only scattering a small
fraction of the Ly$\alpha$ photons. We will explore the evolution with
redshift of these properties in a forthcoming paper.

We show the size distribution measured directly on the 2-d spectra in
the bottom panel of Fig.~\ref{size_dist}: due to the poorer seing of
these observations with respect to the CFHTLS ones, this distribution
is peaked at 0.9 arcsec. Hence, this justifies our choice to simulate
lines as gaussians with an average FWHM of 0.9 arcsec to estimate the
slit flux loss in Section \ref{sect:loss}.

\subsection{Spectral properties of the LAE}\label{sect:spectropro}

\begin{small}
\begin{figure}
\centering \includegraphics[width=1.0\linewidth]{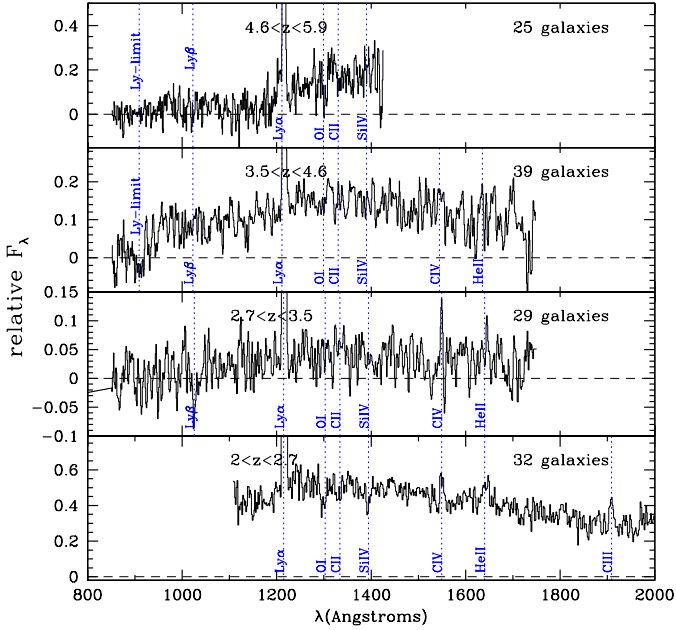}
\caption{Combined spectra of the serendipitous \lae emitters, in
several redshift ranges from z=2 to z=5.9. The displayed range of flux
has been set to show faint continuum features, while the peak of the
Ly$\alpha$ line is largely off-scale. Blue dotted lines indicate
different spectral features depending on the redshift range: Ly-break
at 912\AA~, Ly$\beta$ at 1026\AA~, OI at 1303\AA~, CII at 1333\AA~,
SiIV at 1397\AA~, CIV at 1549\AA~, HeII at 1640\AA~ and CIII at
1909\AA~.  The horizontal dashed lines indicate the level of zero
flux.}
\label{all_lya}
\end{figure}
\end{small}

\begin{small}
\begin{figure}
\centering \includegraphics[width=1.0\linewidth]{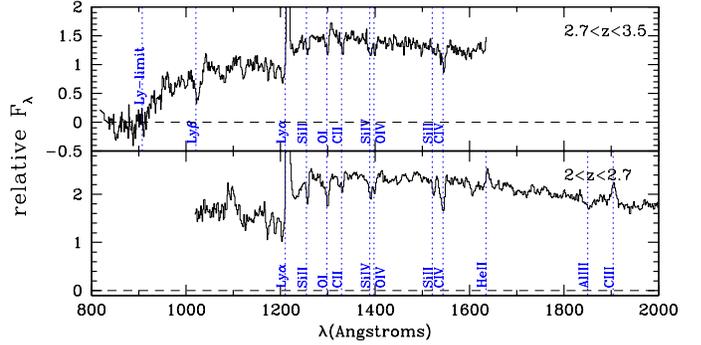}
\caption{Combined spectra of the target \lae emitters in the redshift
ranges $2 \leq z \leq 2.7$ and $2.7 \leq z \leq 3.5$.  The displayed
range of flux has been set to show faint continuum features, while the
peak of the Ly$\alpha$ line is largely off-scale.  The horizontal
dashed lines indicate the level of zero flux.}
\label{all_target}
\end{figure}
\end{small}

We produced combined spectra for all the \lae in our sample, including
primary VVDS targets as well as sources identified serendipitously.
In order to preserve line shapes and to avoid averaging out faint
features, we used accurate redshift measurements from a half-gaussian
fit to the red wing of the Ly$\alpha$ lines, that has been used to
determine the central wavelength of Ly$\alpha$, hence the redshift.
Each spectrum has then been de-redshifted to rest-frame, normalized
using the continuum flux in 1400--1800\AA~ up to z=4.6, 1400--1600\AA~
for $4.6 \leq z \leq 5.9$, and the Ly$\alpha$ line itself for $5.9 < z
\leq 6.7$, then averaged using a sigma-clipping algorithm.  The
results in several redshift ranges are shown in Figure \ref{all_lya}
and Figure \ref{all_target}, for serendipitous and target objects,
respectively.  A number of spectral features can be readily identified
at all z, including of course Ly$\alpha$, but also prominent CIV in
absorption as well as HeII in emission well detected in the two lowest
redshift bins.

Bluer than Ly$\alpha$ we can see that the continuum is more absorbed
as redshift increases, with a ratio $(f[1250-1350\AA]/f[1100-1180\AA]$
for serendipitous LAE of 1.34, 1.47, 2.0, 9.3 for redshifts $z\sim
2.3, 3, 4, 5.3$ respectively, and for target LAE a ratio of 1.41 and
1.51 for redshifts $z\sim 2.3, 3$ respectively. For our highest
redshift objects $5.9 < z < 6.7$, uncertainties on the continuum
blueward of Ly$\alpha$ only enables to state that this ratio is higher
than 5. These measurements are comparable to the expected average
extinction by the intergalactic medium, as Madau~et~al. (1995) predict
ratios of 1.3, 1.5, 2.5 and 8.2 for $z\sim 2.3, 3, 4, 5.3$,
respectively. This strengthens our identification of these emission
line objects as LAE.

Ly$\beta-1026$\AA~ is readily observed, and the Lyman-break at 912\AA~
is well detected for redshifts $z>2.9$ placing this feature in our
observed wavelength range, as expected.

To check for a difference between the targeted magnitude limited
sample and the serendipitous LAE, we produced the combined
spectra for all target LAE in the redshift domains $2 \leq z
\leq 2.7$ and $2.7 < z \leq 3.5$ as shown in Figure \ref{all_target},
while above z=3.5 the serendipitous LAE dominate and their combined
spectra are quasi identical to the spectra shown in Figure
\ref{all_lya}. Many spectral features are clearly detected in
absorption: Ly$\beta-$1026\AA~, OI$-1303$\AA~, SiIV$-1397$\AA~.
Besides Ly$\alpha$, CIV$-1549$\AA~, HeII$-1640$\AA~ and
CIII$-1909$\AA~ are all detected in emission.

The Ly$\alpha$ rest-EW increases from $38\pm1$\AA~ at $z\simeq2.3$ to
$370\pm150$ at $z\simeq6.3$. This trend is probably the result of two
effects: a general trend of increasing star formation at higher
redshifts, and a difference between the observed galaxy populations,
with less (more) luminous galaxies observed at lower (higher)
redshifts. The mean EW of HeII$-1640$ for the full sample is
EW=$1.4\pm0.2$\AA~, $2.0\pm0.1$\AA~ and $4.0\pm2.0$\AA~ at
$z\simeq2.3$, $z\simeq3.1$, and $z\simeq4$ respectively, indicating
the presence of a young stellar population a few $10^7$ years old
(Schaerer~2002).  Considering only the serendipitous \lae, the
EW(HeII) is rising from $3.9\pm0.4$ at $z\simeq2.3$ to $14.5\pm1.5$ at
$z\simeq3.1$. This is clearly indicating that a younger stellar
population is present at the higher redshift. The difference of EW
between the full \lae sample and the serendipitous \lae at
$z\simeq3.1$ is significant, $4.0\pm2.0$ versus $14.5\pm1.5$, possibly
again a consequence of the different populations observed, \lae in the
serendipitous sample corresponding to fainter objects with stronger
star formation or less extinction of the Ly$\alpha$ emitted
photons. This difference at the same redshift could therefore indicate
that continuum--faint \lae contain younger stellar populations than
continuum--bright \lae.  This will be the subject of further studies.

The Ly$\alpha$ line in the combined spectra has an asymmetric profile
with the blue wing truncated and an extended red wing, as shown in
Figure \ref{all_lya_z}.  This is expected as gas outflow and
intergalactic absorption impose a sharp blue cutoff and a broad red
wing. This \lae shape and the clear detection of other expected emission
and absorption features in the combined spectra provides further supporting 
evidence that the bulk of the lines in our sample are \lae rather than [OII].

\subsection{Properties of the 6 galaxies with $5.96 \leq z \leq 6.62$}

The combined spectrum of the 6 galaxies identified with $5.96 < z <
6.62$ is shown in Figure \ref{all_z6_lya}. The continuum below
Ly$\alpha$ is barely detected in this redshift range, consistent with
a strong absorption by the intergalactic medium, while we detect a
continuum at the 2$\sigma$ level above 1215\AA. The line profile shown
in Figure \ref{all_z6_lya} is also asymmetric in this redshift range,
reinforcing the identification of the emission lines with Ly$\alpha$.

The distribution of rest EW(Ly$\alpha$) is in the range $70 < EW <
500$, with a mean EW=310\AA~ and a dispersion of 150\AA~.
Malhotra\&Rhoads~2002 suggested that EW$>$300\AA may indicate a top
heavy IMF. However, our measurements are just around this limit, and
moreover very uncertain, so we cannot draw any conclusions on this
issue. Moreover, using Eq.~\ref{eq:sfr} to convert luminosities to SFR
for these 6 galaxies we obtain SFR in the range $4<SFR<22 M_{\odot}
yr^{-1}$, values that are not so extreme as to require a top-heavy
IMF.

None of these galaxies are photometrically detected, neither in the
single bands reaching as faint as i$_{AB}=28$ ($1\sigma$), nor in the
combined ugirz image, reaching an equivalent magnitude of $AB\simeq
29$.

\subsection{Photometric properties}
\begin{figure}
   \centering\includegraphics[width=1.0\linewidth]{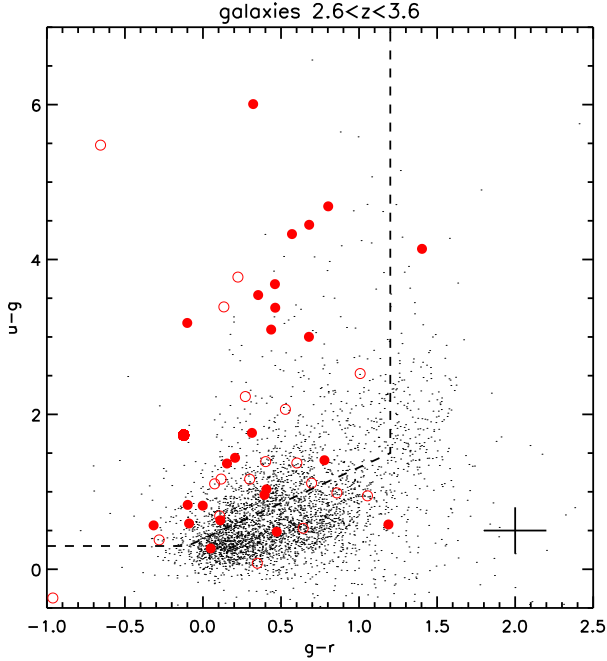}
   \caption{UGR color-color plot for galaxies in our sample between
$2.6<z<3.6$. Filled and open circles represent respectively targets
and serendipitous objects with optical counterpart. Black points show
galaxies at all redshifts. The dashed box indicates the drop-out
region in which galaxies at such redshifts are expected to lie. The cross
in the bottom right part of the diagram shows the typical error bars.}
   \label{UGR}%
    \end{figure}

\begin{figure}
   \centering\includegraphics[width=1.0\linewidth]{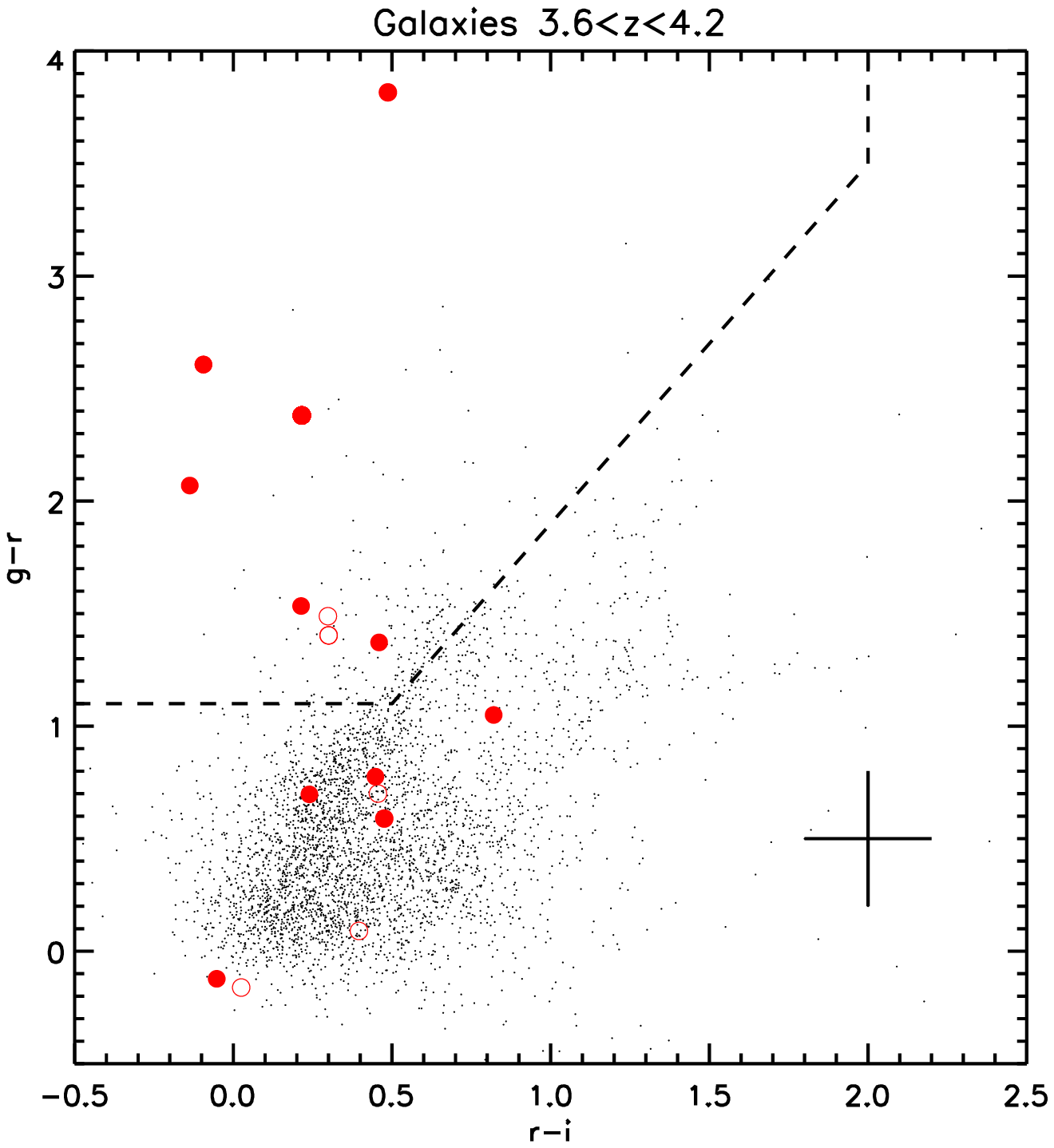}
   \caption{GRI color-color plot for galaxies in our sample between
$3.6<z<4.2$. Filled and open circles represent respectively targets
and serendipitous objects with optical counterpart. Black points show
galaxies at all redshifts. The dashed box indicates the drop-out
region in which galaxies at such redshifts are expected to lie. The cross
in the bottom right part of the diagram shows the typical error bars.}
   \label{GRI}%
    \end{figure}

In Figures~\ref{UGR} and \ref{GRI} we show the ugr and gri diagrams
for all galaxies in our sample (serendipitous and targets) at
$2.6<z<4.2$. In fact, young star forming galaxies usually display very
red colors around the 912\AA~Lyman limit: the flux below 912\AA~is
virtually set to zero by the IGM absorption, while the star formation
activity produces strong emission beyond Ly$\alpha$ line, producing a
very pronounced and high S/N feature. This peculiarity is extensively
used to look for high redshift ($z>2$) galaxies
(e.g. Steidel~et~al.~1999; Giavalisco~2002;
Bouwens~et~al.~2007; 2009; 2010). We determined the part of the
color-color plots in which high-z ``dropouts'' are expected to be
convolving synthetic spectral energy distributions for star-forming
galaxies with the SDSS photometric system filters.

In Fig.~\ref{UGR}, about 80\% of the galaxies with $2.6<z<3.6$ fall in
the expected u-band dropout region. Similarly, in Figure~\ref{GRI},
about 50\% of the galaxies at $3.6<z<4.2$ fall in the expected g-band
dropout region. The fact that a significant fraction of objects with
secure spectroscopic redshifts falls out of the predicted color
selection boxes has already been noted by Le F\`evre~et~al. (2005b),
also for LAE as Gronwall~et~al. (2007) find a significant fraction of
their sample outside of a UVR LBG box, and is mainly the result of the
photometric measurement process (Le F\`evre~et~al., 2010, in prep.).

   \begin{figure*}
   \centering\includegraphics[width=1\linewidth]{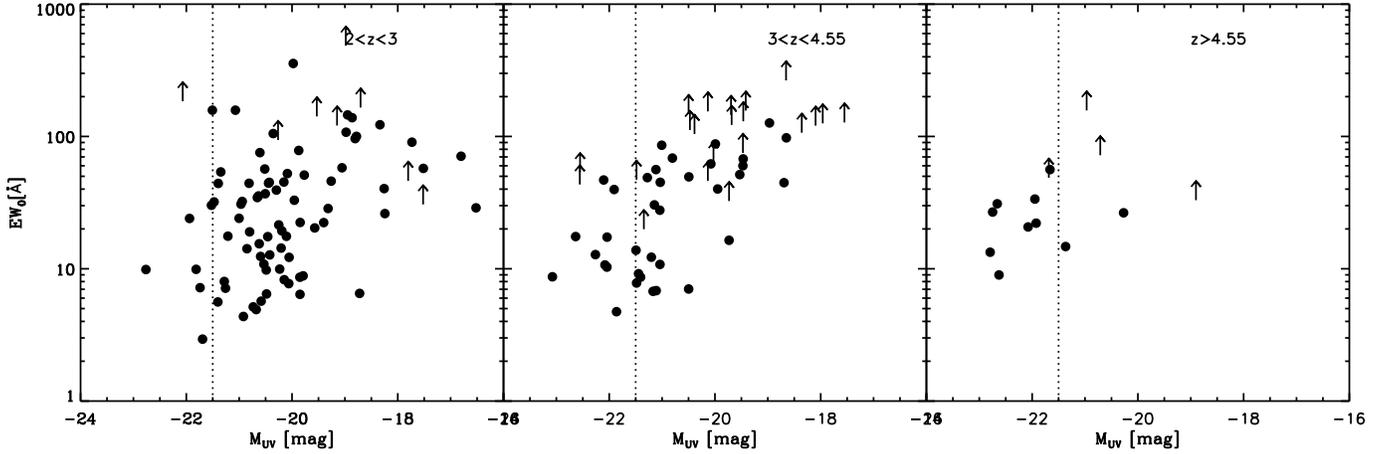}
   \caption{Equivalent Width of the \lae line as a function of the
   absolute magnitude at 1600\AA, for the objects with photometric
   counterpart,in three different redshift bins. Objects with detected
   continuum are shown as filled dots. Objects without a detected
   continuum, for which the equivalent width is only a lower limit,
   are shown as upperward arrows. The dotted line shows
   $M_{UV}=-21.5$: objects brighter than this limit show a
   deficit of large equivalent widths (EW$_0>$20).}
   \label{photo}%
    \end{figure*}

In Figure~\ref{photo} we plot, for the objects with a photometric
counterpart (primary targets and serendipitous), the equivalent width
of the \lae lines as a function of the absolute magnitude in the UV at
$\lambda_{rest}$=1600\AA, for three different redshift
bins. Shimasaku~et~al.~(2006), Ando~et~al.~(2006) and
Ouchi~et~al.~(2008) suggested that objects with $M_{UV}<-21.5$
show typically low equivalent width ($EW<20$), while objects with
$M_{UV}>-21.5$ span a wide range of EW ($50<EW<150$). Our data are
consistent with these findings, even though some of our galaxies
with $M_{UV}<-21.5$ have just a lower limit to the EW. 
We note also that many of the serendipitous galaxies do not appear in
this Figure, as they have no photometric counterparts.

   \begin{figure}
   \centering\includegraphics[width=8cm]{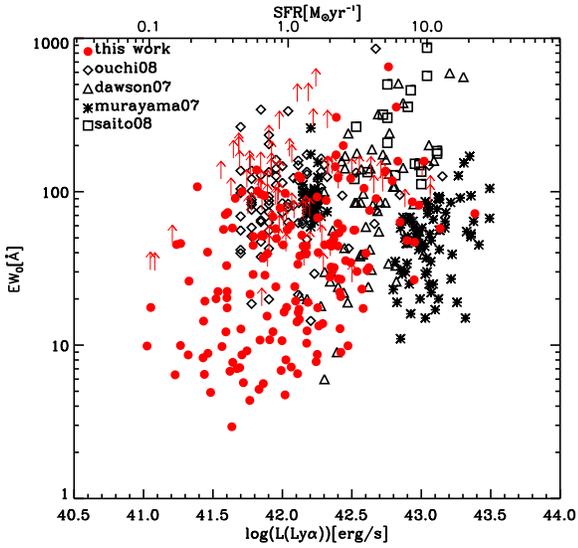}
   \caption{ Luminosity versus equivalent width for the galaxies in
our sample ({\it red points}); filled and open circles indicate
respectively galaxies with and without a detected continuum. For the
latter, the measure of the EW is just a lower limit. Black diamonds,
triangles, crosses and squares show respectively the samples of
Ouchi~et~al.~(2008), Dawson~et~al.~(2007), Murayama~et~al.~(2007) and
Saito~et~al.~(2008). }
   \label{lum_ew}%
    \end{figure}

In Figure~\ref{lum_ew} we compare the \lae Luminosity, the \lae
Equivalent Width and the star formation rate inferred from the \lae
Luminosity for galaxies in our sample with others in literature.
The aim of this figure is to show the different selection criteria for
the various samples of \lae emitters. Narrow band surveys typically
select objects with $EW_0>$20 and do not reach \lae emitters less
luminous than $L_{Ly\alpha}=10^{41.5}$, while our sample has no limits
on the equivalent width and goes 0.5--1 dex deeper in luminosity.
Overall, all the samples seem to follow a broad relation between \lae
luminosity and \lae equivalent width.  Moreover, it is clear that the
typical object in our sample has a star formation rate SFR$\sim1
M_{\odot} yr^{-1}$, with many having even smaller rates, while other
samples have typically SFR$>1 M_{\odot} yr^{-1}$, as expected from our
deeper flux limit.

\section{Luminosity functions}

\subsection{Formalism}

We aim to combine the serendipitous and target \lae emitters in the
U-DEEP and DEEP surveys to produce luminosity functions at different
redshifts. Since the targets and serendipitous objects have very
different selection criteria, for the luminosity function calculation
we treated them independently. So, each galaxy from the target sample
has a weight calculated with respect to the photometric selection, and
each galaxy from the serendipitous sample has a weight calculated with
respect to the spectroscopic selection.

Each of the two sample, however, is the combination of subsamples
having different magnitude or \lae flux limits and covering different
areas: serendipitous LAE have been identified because their line flux
exceeds the spectroscopic flux limit; those coming from the Ultra-Deep
have a flux limit $\sim1.5\times10^{-18} erg/cm^2/$, while those
coming from the Deep survey have a flux limit of $\sim5\times10^{-18}
erg/cm^2/s$. The VVDS primary spectroscopic targets are selected
according to their magnitude in the $I$-band; those coming from the
Ultra-Deep are selected to have $m_I<24.75$, and those coming from the
Deep to have $m_I<24$.

To combine together subsamples reaching different flux (or magnitude)
limits in different areas, we used the extended version of the
$1/V_{max}$ formalism developed by Avni\&Bahcall~(1980). For each
object we computed two effective volumes $V_{max}(j)$, one for the
ultradeep limit and the other for the deep one (if the object is a
target, the limit is the magnitude limit, if the object is a
serendipitous one, the limit is the \lae flux limit). So, for an
object with redshift $z$ assigned to the redshift bin $z_1<z<z_2$:
\begin{equation}
V_{max}(j)=\theta(j)\int_{z_1}^{z_{sup}(j)} \frac{dV}{dz}dz
\end{equation}
where $z_{sup}$ is the minimum between $z_2$ and the redshift at which
the object could have been observed within the limits of the $j$th
selection, $\theta$ is the solid angle covered by the $j$th survey,
and $dV/dz$ is the comoving volume element. For each object, $V_{max}$
is defined as:
\begin{equation}\label{equ:vmax}
V_{max}=\sum_jV_{max}(j).
\end{equation}
This basically means that the brightest objects, that are visible in
both the ultradeep and deep surveys, are weighted according to the
effective volume accessible to them in both surveys, while the
faintest one, that are visible just in the ultradeep survey are
weighted just with respect to this latter. The target sampling rate
and the success rate have been taken into account in computing the
volumes for the targets. They are $\sim0.1$ and $\sim0.3$ for
Ultradeep and Deep surveys, respectively.

The \lae luminosities have been here corrected for the flux slit loss,
as explained in Sect.~\ref{sect:loss}: for the targets, the expected
slit loss is 15\%, and for the serendipitous sample is $\sim$45\%. We
verified that the choice of an average value for such correction do
not affect our main results: in particular, the Schechter best-fit
values presented in Section~\ref{sect:lf} do not change, within the
errors, if the slit flux loss correction are randomly extracted for
each object from the distribution shown in
Fig.~\ref{slit_loss}. Finally, we combined the two $V_{max}$ for them
using Equation~\ref{equ:vmax}.  Moreover, a completeness correction
has been applied to each object according to its flux and redshift,
both for targets and for serendipitous emitters.

{Once we have $V_{max}$ for both the targets and serendipitous
 objects, we compute the galaxy number density for each
 $\Delta\log(L)$ and $\Delta z$ bin as:
\begin{equation}
\phi(L,z)=\frac{1}{\Delta\log(L)}\sum_n\frac{1}{V_{max}}
\end{equation}
where $n$ is the number of objects in that bin.

\subsection{Evolution of the luminosity functions}\label{sect:lf}

We estimated the luminosity functions for 3 redshift bins:
$z=1.95-3.$, $z=3.-4.55$, $z=4.55-6.6$ in order to keep enough
galaxies per redshift bin for a reliable estimate of the LF. The
results are shown in Figure \ref{lumfunc}.

   \begin{figure}
   \centering\includegraphics[width=1.0\linewidth]{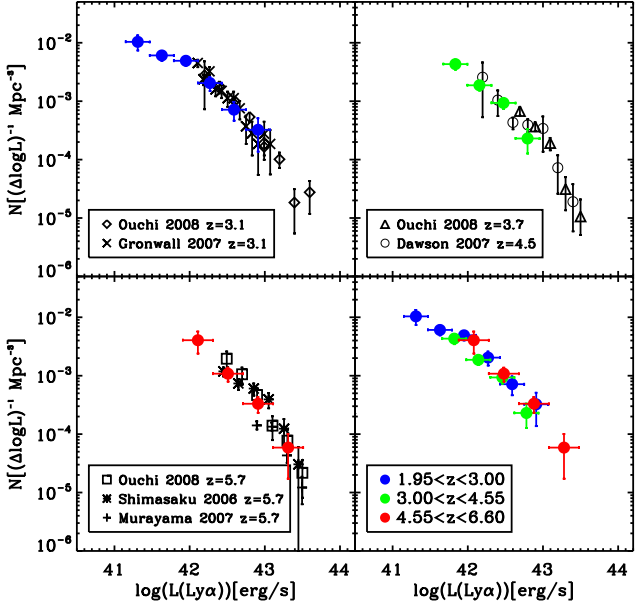}
   \caption{LAE Luminosity function in different redshift bins. No IGM
absorption correction has been applied here. Upper left, upper right
and bottom left panels indicate $2<z<3.2$, $3.2<z<4.55$ and
$4.55<z<6.6$ respectively. The error bars reflect the poissonian
errors. Black datapoints show literature data, spanning from z=3 to
z=5.7: diamonds, triangles and squares represent Ouchi~et~al.~(2008)
at respectively $z=3.1$, $z=3.7$ and $z=5.7$; asterisks, plus signs,
crosses and circles indicate respectively Shimasaku~et~al.~(2006) at
$z=5.7$, Murayama~et~al.(2007) at $z=5.7$, Gronwall~et~al.~(2007) at
$z=3.1$ and Dawson~et~al.~(2007) at $z=4.5$. We group together our
measurements in the different redshift bins in the bottom right panel,
to show that within the error bars, the \lae LF does not evolve from
$z=2.5$ to $z=6$.}
   \label{lumfunc}%
    \end{figure}

When comparing our datapoints in these different redshift intervals,
we can see that the observed apparent luminosity function (i.e. the
non-IGM corrected LF) does not evolve in the redshift range
$z\simeq2.5$ to $z\simeq6$ within our error bars. Although our data
cover a slightly wider redshift interval they are in agreement with
the literature data, within the error bars.  This is quantified below.

The unprecedented depth of our survey enables to extend the luminosity
function at $z=2-3.$ down to $\log (L_{Ly\alpha})$=41.3, about one
order of magnitude deeper than literature data at similar redshifts.
This allows us to strongly constrain the slope of the luminosity
function at $z\sim$2.5. This parameter is extremely important, because
a small change of this slope can produce a large change in the
luminosity density.

We assumed here that the luminosity function of \lae emitters is well
represented by a Schechter law (Schechter~1976):
\begin{equation}
\Phi(L)dL=\Phi^*(L/L^*)^{\alpha}\exp(-L/L^*)d(L/L^*)
\end{equation}

We obtained the best-fit functions for the 3 redshift bins using this
Schechter function.  As a first try, we fitted the luminosity
functions in the 3 redshift intervals allowing all the 3 parameters
describing the Schechter function to vary. However, by doing this, the
typical luminosity $L^*$ in the first two bins and the slope $\alpha$
in the high redshift bin are poorly constrained. Indeed, in the first
two bins we do not have \lae galaxies brighter than $10^{43}
erg/cm^2/s$, and in the last one we do not observe emission lines
fainter than $10^{42} erg/cm^2/s$. So, we decided to use other
datapoints in the literature to constrain $L^*$ in the first two
bins. In particular, we averaged the $L^*$ values obtained by
Ouchi~et~al.~(2008) and Gronwall~et~al.~(2007) at $z=3.1$, obtaining
$\log(L^*)=42.7$. So, in our fit procedure we fixed
$\log(L_{Ly-\alpha})=42.7$ in the first two bins. We verified that
this approach is equivalent to including Ouchi~et~al.~(2008) and
Gronwall~et~al.~(2007) datapoints with $\log(L_{Ly\alpha})\gtrsim43$ to
ours and performing the fit.  Moreover, we constrained $\alpha$ in the
last bin to the average of the value in the first two bins. The
best-fit parameters are reported in Table \ref{tab:lumfunc}.
\begin{table*}\label{tab:lumfunc}
\label{schechter} \centering
\begin{tabular}{clr@{$\pm$}rr@{$\pm$}lr@{$\pm$}lr@{$\pm$}lr@{$\pm$}lc}
\hline\hline \noalign{\smallskip} \multicolumn{1}{c}{$z$} &
\multicolumn{2}{c}{$z=1.95\div3$} & \multicolumn{2}{c}{$z=3\div4.55$}
& \multicolumn{2}{c}{$z=4.55\div6.6$}\\ \noalign{\smallskip} \hline
\noalign{\smallskip} $\alpha$
&\multicolumn{2}{c}{$-1.6_{-0.12}^{+0.12}$}
&\multicolumn{2}{c}{$-1.78_{-0.12}^{+0.10}$}
&\multicolumn{2}{c}{$-1.69$}\smallskip
\\ 
$\Phi^*$
&\multicolumn{2}{c}{$7.1_{-1.8}^{+2.4}\times10^{-4}$}
&\multicolumn{2}{c}{$4.8_{-0.8}^{+0.8}\times10^{-4}$}
&\multicolumn{2}{c}{$9.2_{-1.9}^{+2.3}\times10^{-4}$} \smallskip\\

$\log(L^*_{uncorr})$ &\multicolumn{2}{c}{$42.70$}
&\multicolumn{2}{c}{$42.70$}
&\multicolumn{2}{c}{$42.72_{-0.12}^{+0.10}$}\smallskip\\ 
$\log(L^*_{corr})$
&\multicolumn{2}{c}{$42.74$}
&\multicolumn{2}{c}{$42.83$} 
&\multicolumn{2}{c}{$43.0_{-0.12}^{+0.10}$}\\ 
\noalign{\smallskip} \hline
\noalign{\smallskip}
\end{tabular}
\caption{Schechter Function Parameters.  $\log(L^*_{uncorr})$ and
$\log(L^*_{corr})$ indicate respectively the value uncorrected and
corrected for the IGM absorption. Error bars give the 1$\sigma$
confidence levels. Values have been fixed during the fitting
procedure when no error bars are shown.}\end{table*}

Importantly, we note that the luminosity functions reported above are
not corrected for absorption by the intergalactic medium. The \lae
flux from high-$z$ sources is generally absorbed by neutral hydrogen
present in the IGM, that absorbes the blue wing of the \lae line,
producing an asymmetric line profile (Hu~et~al.~2004;
Shimasaku~et~al.~2006). We therefore measure IGM absorbed \lae fluxes,
and not {\it intrinsic} \lae fluxes. Moreover, the amount of
absorption is not the same at different redshifts: at high $z$ the
amount of intervening intergalactic medium is larger. As a
consequence, if the {\it observed i.e apparent} luminosity functions
do not evolve from $z\sim2.5$ to $z\sim6$, the {\it intrinsic} one
positively evolves. The IGM optical depths have been estimated in
various studies (i.e. Madau~1995, Fan~et~al.~2006 and Meiksin~2006):
all the authors agree that the amount of absorption increases from
$\sim$15\% at $z\sim$3 to $\sim$50\% at $z\sim$6 (at this redshift,
basically all the blue side of the line is absorbed).  Using the
standard radiative transfer prescription by Fan~et~al.~(2006) for the
IGM optical depths, we can obtain the intrinsic luminosities $L_{int}$
from the observed $L_{obs}$. At $z\sim$2.5, 4.2 and 6 the ratio
$L_{obs}/L_{int}$ is respectively 0.91, 0.73 and 0.52.

   \begin{figure}
   \centering\includegraphics[width=1.0\linewidth]{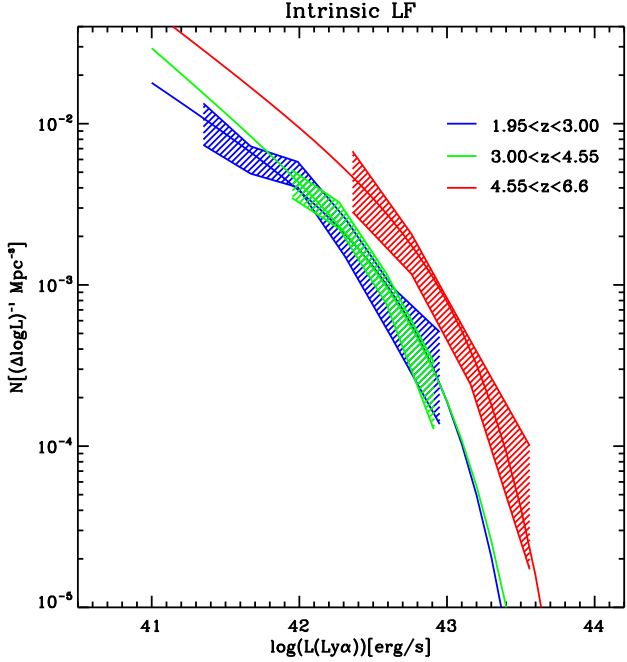}
   \caption{We report here our estimates of the intrinsic luminosity
functions, corrected for the IGM absorption (shaded
regions). Blue, green and red indicate respectively $1.95<z<3$,
$3<z<4.55$ and $4.55<z<6.6$ ranges. Solid lines show our best
fits to the data. }
   \label{lumfuncorr}%
    \end{figure}

In Figure~\ref{lumfuncorr} we show instead the {\it intrinsic} luminosity
functions, corrected for the IGM absorption according to the prescription
of Fan~et~al.~(2006), together with our Schechter fits to the data.

The first main result from this analysis is that there does not seem
to be a strong evolution of the {\it apparent} luminosity function
between $z\sim6$ and $z\sim2.5$, within our error bars. In fact,
looking at Fig.~\ref{lumfunc} we see that within our 1$\sigma$ errors,
the luminosity functions in the different redshift bins overlap;
moreover, looking at the Schechter best fit parameters in Table
\ref{tab:lumfunc}, we can see again that within 1$\sigma$ the
Schechter parameters do not evolve. On the other hand, by looking at
the {\it intrinsic} luminosity functions in Fig.\ref{lumfuncorr}, we
do see an evolution between $z\sim6$ and $z\sim4$, while no sizeable
evolution is observed between $z\sim4$ and $z\sim2$.  The observed
evolution can be parametrized with the evolution in $L^*$, that is at
$z\sim6$ about 1.8 times higher than in the first bin.  However, its
significance level is only 1.5$\sigma$.

The second interesting result comes from our ability to constrain the
faint end of the luminosity function.  We find slopes
$\alpha=-1.6_{-0.12}^{+0.12}$ at $z\sim$2.5 and
$-1.78_{-0.12}^{+0.10}$ at and $z\sim$4. Our data formally exclude a
flat slope $\alpha \sim -1$ at 5 and 6.5$\sigma$, at these two
redshifts.  Our slope values are significantly better constrainted
than the best value of 1.36 and the marginalized value of
$-1.49^{+0.45}_{-0.34}$ found at $z\sim3.1$ by Gronwall~et~al.~(2007),
a consequence of the 10 times fainter flux limit of our sample.  Note
that other authors, who again do not have deep enough data, do not try
to fit $\alpha$, but they rather fix it to some plausible value (-1,
-1.5 and -2 for Ouchi~et~al.~2008; -1.6 for Dawson~et~al.~2007; -1.2
and -1.6 for Lemaux~et~al.~2009). Our analysis therefore establishes
the first reliable estimate of the faint end slope of the luminosity
function of \lae galaxies at $z<5$.

It is also important to state that the possible ``non-Ly$\alpha$'' emitters
in our sample, described in Sect.~\ref{lae:id}, do not strongly affect
these results. First of all, the luminosity distribution of the
``ambiguous'' lines (the 49 lines that cannot be unambiguously
identified as \lae based on the line ratios) is similar to that of the
global sample, and thus it is not expected to affect the slope
determination. However, it is possible that in the $4.55<z<6.6$
redshift bin the contamination is higher than in the others. In fact,
for all the lines at $\lambda>6920$ (corresponding to
z$(Ly\alpha)\gtrsim4.7$), we cannot check for other lines in the
spectra, if the line is identified as [OII] (see Sect.\ref{lae:id}).
However, we showed that the contamination is not expected to be
higher than 10\%, based on the EW distribution. Thus, assuming that
all the 17 lines that can be identified as [OII] are at $z>4.7$ would
decrease the luminosity function in the last bin of a factor of 50\%
at the most. This would imply a weaker evolution of the luminosity
function from $z=6$ to $z=4$.

\subsection{Evolution of the star formation rate density}

With these new constraints on the evolution of the \lae luminosity
function at these redshifts, it is interesting to estimate the
contribution of the \lae emitters to the global star formation rate
density of the Universe.  This is not trivial, as \lae emission
produced not only by star formation activity, but also by other
processes like cooling radiation, AGN activity or shock
winds. Moreover, \lae emission is attenuated by IGM and dust.

We are computing here the SFRD using only the {\it intrinsic}  \lae luminosity
functions. In order to estimate the contribution of the galaxies in
our sample, we integrated the luminosity function from
$L_{Ly\alpha}=0.04\times L*$ to $\log(L_{Ly\alpha})$=44 (roughly the
interval covered by our data). We then converted these \lae luminosity
densities in star formation rate densities by using Equation~\ref{eq:sfr}.

   \begin{figure}
   \centering\includegraphics[width=1.0\linewidth]{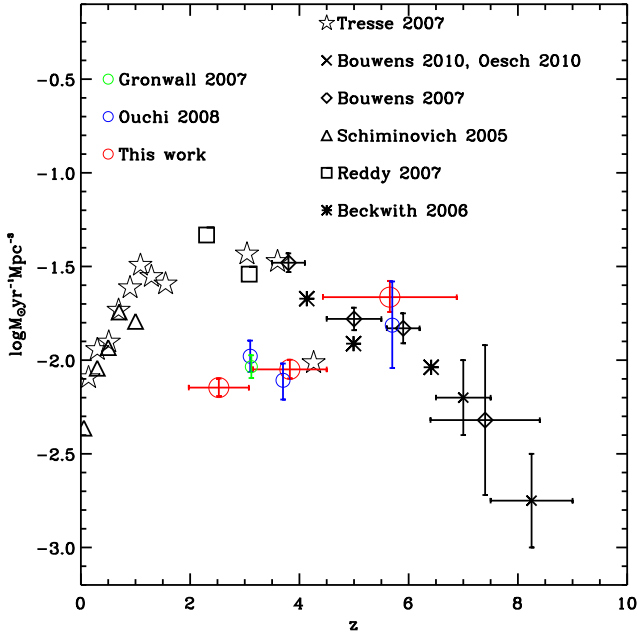}
   \caption{Evolution of the star formation rate density as a function
of the redshift, inferred integrating the intrinsic \lae luminosity
functions down to $0.04\times L*$. Red open circles are our data,
while blue and green open circles represent other \lae SFRD
respectively by Ouchi~et~al.~(2008) and Gronwall~et~al.~(2007). ,
Open triangles, open squares, open lozenges, stars and crosses are UV
estimates respectively by Schiminovich~et~al.~(2005),
Reddy~et~al.~(2008), Bouwens~et~al.~(2007), Beckwith~et~al.~(2006),
Tresse~et~al.~(2007) and Bouwens~et~al.~(2010) }

\label{sfd}%
\end{figure}

The derived star formation rates are shown in Figure~\ref{sfd},
together with the most recent estimates of the SFRD at redshift between
$z$=0 and $z$=6. Ouchi~et~al~(2008) and Gronwall~et~al.~(2007) use a
sample of narrow band selected \lae emitters, while the other
estimates are based on the UV galaxy luminosity function. 

We can see that our measurements show a slight evolution of the \lae
SFRD between $z\sim$2.5 and $z\sim$4, and a much more significant
evolution from $z\sim4$ and $z\sim$6. Our values are compatible within
the errors to those obtained by Ouchi~et~al.~(2008).

When comparing our estimates of the \lae SFRD with the UV-derived
ones, we see that the total contribution of \lae galaxies to the
global star formation density at $z\sim2-6$ is important, increasing
from $\sim20\%$ at $z\sim$2.5 to $\sim100\%$ at $z\sim6$.  This
implies that the \lae emission is a good tracer of the star formation,
this statistically implies that, while at redshift $z\sim2$ only a
small fraction of the galaxies contributing to the star formation
history of the universe also show a \lae emission, all the galaxies at
$z\sim6$ do show \lae in emission. In other words, the so called {\it
escape fraction}, that is the fraction of \lae emission produced by
the star formation that actually escapes the star formation regions
changes a lot with the cosmic epoch, from 20\% at $z=2.5$, reaching up
to 100\% at $z\sim6$. This seems to indicate that the mechanism which
is absorbing the \lae photons in most of the galaxies at $z=2$ is not
effective at $z\sim6$, as is expected in a very low dust
medium. However, dust estimates at $z\simeq5-6$ show that the dust
content should be sufficient to produce significant Ly$\alpha$ photons
absorption (e.g. Bouwens~et~al.~2009).

Alternatively, it is possible that the UV luminosity function based on
Lyman break galaxies searches is based on more and more incomplete
counts at increasingly high redshifts, and that the current
UV--derived SFRD are underestimates (e.g. Le F\`evre~et~al.~2005b;
Paltani~et~al.~2007). Current estimates of the luminosity density at
$z\simeq5-6$ agree within a factor 2--3 or so
(Bouwens~et~al.,~2007,~2009), and will need new generation surveys to
be improved.

\section{Summary}

In this paper we have reported the discovery of 217 faint LAE in the
range $2 \leq z \leq 6.62$ from targeted and serendipitous very deep
observations using the VIMOS multi-slit spectrograph on the VLT.
Adding together the areas covered by each slitlet combined to the
wide wavelength coverage 5500-9350\AA~ of the Deep survey and
3600-9350\AA~ for the UltraDeep, we surveyed effective sky areas of
22.2 $arcmin^2$ and 3.3 $arcmin^2$ respectively. This produces a
survey volume of $\sim2.5\times10^5 Mpc^3$, observed to unprecedented
depth F$\sim 1.5\times10^{-18}erg/s/cm^{2}$.  This volume is about one
order of magnitude bigger than all other spectroscopic surveys
produced up to now at comparable fluxes:
van~Breukelen,~Jarvis~\&~Venemans~(2005) sampled $10^4 Mpc^3$ down to
1.4$\times10^{-17} erg/cm^2/s$; Martin~et~al.~(2008) sampled
$4.5\times10^4 Mpc^3$ down to similar fluxes in a narrow redshift
range. Narrow band imaging surveys sampled bigger volumes than ours,
but to a shallower flux: Ouchi~et~al.~(2008) covered a volume of
$\sim10^6 Mpc^3$, down to fluxes $\sim2\times10^{-17} erg/cm^2/s$.
Serendipitous surveys have been presenting low number of objects
(Sawicki~et~al., 2008), even if somewhat deeper (Malhotra~et~al.,
2005; Rauch~et~al., 2008).  We are therefore sampling deeper into the
LAE luminosity function as we discussed in section \ref{sect:lf}.

From an observational point of view, we demonstrate the efficiency of
blind LAE searches with efficient multi-slit spectrographs. The
success of our approach is the result of combining a broad wavelength
coverage to a large effective sky area, with long integration times,
made possible by the high multiplex of the VIMOS instrument. The broad
wavelength coverage has been essential to secure the spectroscopic
redshifts from one single observation, without the need for follow-up
to confirm the \lae nature of the emission lines detected. This
observing efficiency compares favorably with the time needed to
perform narrow band imaging searches followed by multi-slit
spectroscopy, and comparing the wide range in redshift covered by the
former versus a narrow range for the latter.  When the density of
faint LAE is high, of the order several LAE/arcmin$^{2}$, multi-slit
spectrographs become more efficient to secure a large number of
confirmed sources than narrow band imaging searches, while at bright
fluxes covering a wide field is essential to find rarer sources and
narrow band imaging is more efficient.  The two approaches will
therefore remain complementary.

Our main findings are the following:
   \begin{enumerate}
   \item We found a total of 217 LAE with confirmed spectroscopic
     redshifts in the range $2 \leq z \leq 6.62$, 133 coming from the
     serendipitous discovery in the multi-object spectrograph slits of
     the VVDS (105 from the Ultra-Deep and 28 from the Deep), and 84
     coming from targeted VVDS observations of galaxies with $17.5
     \leq i_{AB} \leq 24.75$ (Le F\`evre~et~al., 2010, in
     prep.). About 50\% of the Ultra-Deep and 40\% of the Deep
     serendipitous targets have a detected optical counterpart down to
     magnitude $AB\sim28$ in deep CFHTLS images.
   \item 
  The observed projected density of LAE with a \lae emission brighter
     than F$\sim 1.5\times10^{-18}erg/s/cm^{2}$ in the range $2 \leq z
     \leq 6.6$ is $33$ LAE per arcmin$^2$, with $25$ LAE per
     arcmin$^2$ with $2 \leq z \leq 4.5$ and $8$ LAE per arcmin$^2$
     with $4.5 < z \leq 6.62$. The corresponding volume density of
     faint LAE with $L(Ly\alpha) \geq 10^{41}$ergs.s$^{-1}$ is $\sim4
     \times 10^{-2}$Mpc$^{-3}$, a high density not yet observed at
     these redshifts.
   \item
     The mean rest-frame EW(\lae) of LAE in our sample range from
     about 40\AA~ at $z\sim2-3$ to $\sim300-400$ at z$\sim5-6$, and
     the star formation rate covers a wide range
     $0.1-20$M$_{\odot}$yr$^{-1}$, assuming no ISM or IGM extinction
     and a Salpeter IMF. The HeII-1640\AA~ emission has EW$\sim4-14$
     at $z\sim2-4$ indicating the presence of young stars of a few
     Myr.  We therefore detected vigorously star forming galaxies as
     well as galaxies with star formation comparable to dwarf
     starburst galaxies at low redshifts.
   \item
     The \lae {\it apparent} luminosity function does not evolve
     between z=2 and z=6, within the error bars of our survey. Taking
     into account the average differential evolution in the IGM
     absorption with redshift therefore translates into a positive
     evolution of the {\it intrinsic} \lae LF of about 0.5 magnitude
     from $z\sim2-3$ to $z\sim5-6$.
   \item
     We obtain a robust estimate of the faint end slope of the LAE
     luminosity function from a large sample of spectroscopically
     confirmed LAE.  It is very steep: we find $\alpha\simeq-1.6$ at
     $z\sim$2.5 and $1.8$ at $z\sim$4.
   \item
     The SFRD contributed by \lae galaxies is increasing from
     $5\simeq10^{-3}$M$_{\odot}$yr$^{-1}$Mpc$^{-1}$ at $z\simeq2.5$
     to $\simeq 2\times10^{-2}$M$_{\odot}$yr$^{-1}$Mpc$^{-1}$  at $z\simeq6$.
     The contribution of the \lae galaxies to the total SFRD of the
     universe as inferred by UV luminosity functions reported in the
     literature increases from $\sim20$\% at z=2.5 to $\sim100$\% at
     z=6. This seems to imply that all the galaxies that are forming
     stars at $z=6$ must show \lae emission, therefore are in a very
     low dust medium.  At z=2.5 80\% of the star forming galaxies must
     have the \lae emission produced by star formation blocked by some
     mechanism so that only 20\% of the star forming galaxies show
     \lae emission. A direct consequence would be that the \lae escape
     fraction varies from 0.2 at $z\simeq2.5$ to about 1 at
     $z\simeq6$.  This result would remain robust only if the total
     SFRD estimates based on UV luminosity functions using Lyman break
     galaxies identifications are complete.
   \end{enumerate}

The new VVDS measurements reported here bring a new important
constraint to the LAE LF with a steep faint end slope observed at
$2<z<4.5$.  This implies that LAE with star formation rates of a few
$10^{-1}$M$_{\odot}$yr$^{-1}$ comparable to that of low redshift dwarf
star forming galaxies are dominating the LAE SFRD at these redshifts.  
While it remains to be proven that the faint end slope of the LF stays steep 
going to still higher redshifts, assuming the slope $\alpha \simeq 1.7$ as
implied by our measurements at $2 < z < 4.5$ would imply that the LAE
population is becoming the dominant source of star formation producing
ionizing photons in the early universe $z\sim>5-6$, becoming
equivalent to that derived from Lyman Break Galaxies searches. The
steep faint end slope further implies that during reionisation
sub-L$_*$ galaxies may have played an important role in keeping the
universe ionized.  

These results further demonstrate that efforts
dedicated to constraining the evolution of the luminosity function of
high redshift LAE will remain an important tool to probe into the
reionisation period.

   \begin{acknowledgements}
The authors acknowledge the anonymous referee for the constructive
comments provided. PC acknowledges Sara Salimbeni for the useful discussions.
   \end{acknowledgements}

\end{document}